\documentclass{article}

\usepackage{microtype}
\usepackage{graphicx}
\usepackage{subcaption}
\usepackage{booktabs} 

\usepackage{hyperref}

\usepackage[preprint]{icml2026}

\usepackage{amsmath}
\usepackage{amssymb}
\usepackage{mathtools}
\usepackage{amsthm}

\usepackage[capitalize,noabbrev]{cleveref}

\theoremstyle{plain}

\theoremstyle{definition}

\theoremstyle{remark}

\usepackage[textsize=tiny]{todonotes}

\usepackage[utf8]{inputenc}
\usepackage[T1]{fontenc}

\usepackage{caption}
\usepackage{inconsolata}
\usepackage{amssymb,amsmath,amsfonts}
\usepackage{soul}
\usepackage{xspace}
\usepackage{relsize}
\usepackage{listings}
\usepackage{enumitem}
\usepackage{sepnum}
\usepackage{multirow}
\usepackage{multicol}
\usepackage{makecell}
\usepackage{colortbl}
\usepackage[normalem]{ulem}
\usepackage{tikz}
\usetikzlibrary{shapes.geometric,shapes.misc,arrows.meta,matrix,fit,calc,backgrounds}
\usepackage{pgfplots}
\usepgfplotslibrary{statistics}
\pgfplotsset{compat=newest}

\definecolor{ColorLstHighlight}{RGB}{209,214,250}
\definecolor{ColorCodeBackground}{RGB}{250,250,250}
\definecolor{ColorCodeKeyword}{RGB}{0,0,255}
\definecolor{ColorCodeComment}{rgb}{0,0.6,0}
\definecolor{ColorCodeString}{rgb}{0.58,0,0.82}
\definecolor{ColorCodeNumber}{rgb}{0.5,0.5,0.5}
\definecolor{ColorAlgoKeyword}{RGB}{0,0,255}
\definecolor{ColorAlgoComment}{rgb}{0,0.6,0}
\definecolor{ColorAlgoNumber}{rgb}{0.5,0.5,0.5}
\definecolor{ColorDiffAddition}{rgb}{0,0.6,0}
\definecolor{ColorDiffDeletion}{RGB}{255,25,36}
\definecolor[named]{ACMBlue}{cmyk}{1,0.1,0,0.1}
\definecolor[named]{ACMYellow}{cmyk}{0,0.16,1,0}
\definecolor[named]{ACMOrange}{cmyk}{0,0.42,1,0.01}
\definecolor[named]{ACMRed}{cmyk}{0,0.90,0.86,0}
\definecolor[named]{ACMLightBlue}{cmyk}{0.49,0.01,0,0}
\definecolor[named]{ACMGreen}{cmyk}{0.20,0,1,0.19}
\definecolor[named]{ACMPurple}{cmyk}{0.55,1,0,0.15}
\definecolor[named]{ACMDarkBlue}{cmyk}{1,0.58,0,0.21}
\hypersetup{
  colorlinks,
  allcolors=.,
  linkcolor=ACMPurple,
  citecolor=ACMPurple,
  urlcolor=ACMDarkBlue,
  filecolor=ACMDarkBlue
}

\setlist[itemize]{leftmargin=1.5em,nosep}
\setlist[enumerate]{leftmargin=1.5em,nosep}

\newcommand{\codefontsize}{\defaultcodefontsize}
\lstdefinestyle{defaultlststyle}{
  backgroundcolor=\color{ColorCodeBackground},
  basicstyle=\ttfamily\codefontsize,
  keywordstyle=\color{ColorCodeKeyword}\bf\ttfamily,
  commentstyle=\color{ColorCodeComment},
  numberstyle=\tiny\color{ColorCodeNumber},
  stringstyle=\color{ColorCodeString},
  lineskip=0pt,
  breakatwhitespace=false,
  columns=fullflexible,
  keepspaces=true,
  showspaces=false,
  showstringspaces=false,
  breaklines=true,
  breakatwhitespace=true,
  captionpos=b,
  numbers=left,
  numbersep=4pt,
  xleftmargin=0pt,
  frame=tb,
  framerule=0.5pt,
  basewidth=0.5em,
  escapechar=§,
  showtabs=false,
  tabsize=2,
  language=python,
  morekeywords={codoku,cast\_int,global\_chksum,check\_chksum,\_rd,g\_\_chk,v\_\_chk}
}
\newcommand{\code}[1]{\texttt{\smaller[0.5]{#1}}}

\usepackage{lstlinebgrd}
\usepackage{pgffor}

\newcommand{\highlightmultilines}[1]{%
  \lstset{linebackgroundcolor={%
    \xdef\lstHL{\noexpand\color{ColorCodeBackground}}%
    \foreach \x in {#1}{%
      \ifnum\value{lstnumber}=\x%
        \xdef\lstHL{\noexpand\color{ColorDiffAddition!12!white}}%
      \fi%
      \ifnum\value{lstnumber}=-\x%
        \xdef\lstHL{\noexpand\color{ColorDiffDeletion!12!white}}%
      \fi%
    }%
    \lstHL}}}

\newcommand{\defaultalgofontsize}{\footnotesize}

\newcommand{\codokucell}[1]{\code{\sethlcolor{ColorLstHighlight}\hl{\footnotesize#1}}}

\newcommand{\lstcodokucell}[1]{\sethlcolor{ColorLstHighlight}\hl{#1}}
\newcommand{\lstcodokucellsol}[1]{\sethlcolor{ColorLstHighlight}\hl{#1}}

\newtheoremstyle{mydefinitionstyle} 
  {10pt} 
  {10pt} 
  {\normalfont} 
  {1em} 
  {\bf} 
  {.} 
  {.5em} 
  {} 

\theoremstyle{mydefinitionstyle}

\newbool{ShowAppendix}

\newcommand{\ie}[0]{\textit{i.e.}}

\newcommand{\smalltitle}[1]{\noindent\textbf{#1}. }

\begin{document}

\twocolumn[
  \icmltitle{Codoku: Renewable Program-Reasoning Challenges for Frontier Coding Agents}



  \icmlsetsymbol{equal}{*}

  \begin{icmlauthorlist}
    \icmlauthor{Cong Li}{ethz}
    \icmlauthor{Hao Sun}{ethz}
    \icmlauthor{Zenan Li}{ethz}
    \icmlauthor{Zhendong Su}{ethz}
  \end{icmlauthorlist}

  \icmlaffiliation{ethz}{ETH Zurich}

  \icmlcorrespondingauthor{Cong Li}{cong.li@inf.ethz.ch}

  \icmlkeywords{Machine Learning, ICML}

  \vskip 0.3in
]

\printAffiliationsAndNotice{}  

\begin{abstract}
Existing program-reasoning benchmarks ask large language models to predict a program's behavior on a given input.
Coding agents break two assumptions on which these benchmarks rest:
an agent can recover the answer by executing the program instead of reasoning about it,
and fixed task sets drawn from existing programs are increasingly exposed to contamination, yet costly to renew.
We introduce \emph{Codoku} (\emph{code sudoku}), a renewable benchmark in which a solver fills typed cells in a partial program to satisfy global static and dynamic constraints, such as a prescribed control-flow graph and execution path.
Because a partial program cannot be executed and valid fillings are sparse in an exponentially large space of interdependent choices, neither tool use nor enumeration can substitute for program reasoning.
Puzzles are synthesized from scratch via semantic reification, so fresh puzzles of controllable complexity can be generated on demand, each with a witness that guarantees solvability.
We evaluate five frontier models on 300 puzzles through a coding agent free to use any tool within a fixed budget.
Small puzzles already challenge open-weight models, whereas even proprietary models solve only about half of the large ones.
Codoku thus offers a renewable testbed for program reasoning that can keep pace with rapidly improving coding agents.
GitHub: \url{https://github.com/connglli/Codoku}.

\end{abstract}

\section{Introduction}\label{sec:introduction}

The ability to reason about programs, which we refer to as \emph{program reasoning}, is fundamental to large language models (LLMs) on coding tasks.
A substantial line of benchmarks evaluates this ability through the relation among a program, its inputs, and its execution:
given a complete program and a concrete input, an LLM is asked to predict outputs, intermediate states, branch outcomes, execution paths, or unexpected behaviors~\citep{cruxeval,reval,codeio,core,canllmsreason,robustexecution}.
These benchmarks have enabled meaningful comparisons among LLMs and have supported real progress in program reasoning.
They were designed, however, for \emph{bare models}.
Today, LLMs increasingly operate inside \emph{coding agents}, \ie, harnesses that equip them with shells, interpreters, compilers, debuggers, fuzzers, and other tools.
Transplanting these benchmarks into this setting is more than a change of harness:
it invalidates two critical assumptions on which they rest, one concerning their \emph{format} and the other their \emph{supply}.

\emph{Tool access decouples task success from program reasoning}.
Predicting what a program does serves as a proxy for reasoning about it only as long as the predictor cannot execute the program;
an agent, however, can.
Every quantity that these benchmarks measure, such as outputs and intermediate states, is recoverable by execution, instrumentation, or debugging,
so a task can be solved with a few tool calls and without any understanding of the program or its semantics.
Execution-prediction benchmarks such as CruxEval~\citep{cruxeval} and REval~\citep{reval} remain valuable for evaluating bare models, but their task format is no longer suitable for coding agents.

\emph{Nonrenewable benchmarks cannot supply fresh challenges}.
All program-reasoning benchmarks that we are aware of draw their tasks from existing programs: textbook exercises, programming contests, and GitHub repositories.
These tasks are either created by hand for quality or mined automatically for scale.
Both strategies yield valuable tasks, yet both freeze a benchmark at publication time while the LLMs under evaluation keep evolving.
Because the originating programs and their traces are public, each training cycle increases the likelihood that they have been absorbed.
Web access aggravates the problem further, as an agent may simply retrieve the source program or even a published solution.
The natural remedy, renewal, is costly: as LLMs become more capable, renewing these benchmarks at comparable quality and quantity requires substantial effort.

\smalltitle{Codoku: renewable challenges for frontier coding agents}
We introduce a new kind of program-reasoning challenge, \emph{codoku} (short for \emph{code sudoku}).
Like sudoku, a codoku puzzle asks a solver to fill \emph{cells} whose choices are linked by \emph{global constraints}.
Whereas sudoku constrains the digits in each row, column, and block,
codoku constrains a program's static structure and dynamic behavior.

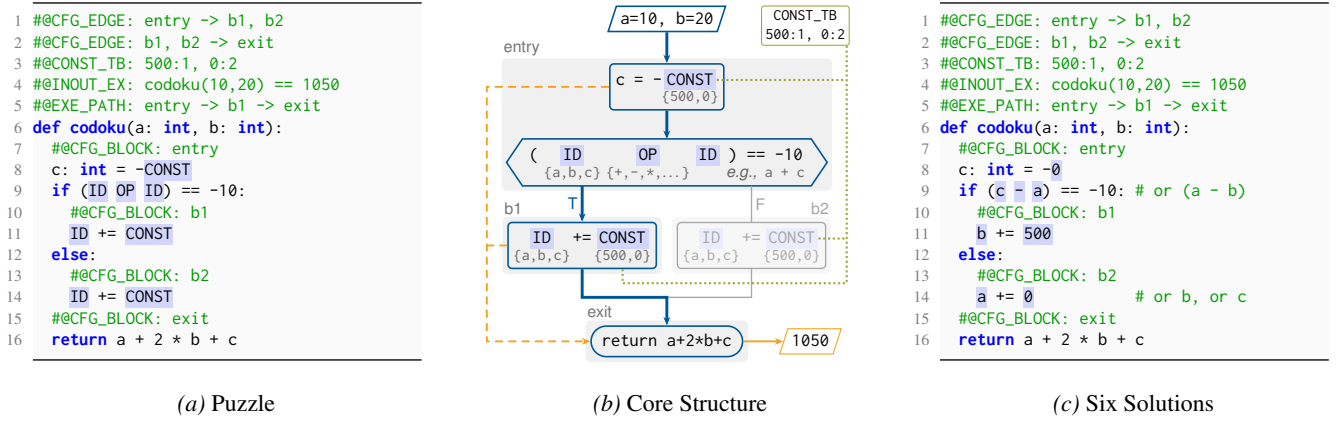
\begin{figure*}[t]
    \renewcommand{\codefontsize}{\scriptsize}
    \colorlet{ColorFigPath}{ACMDarkBlue}
    \colorlet{ColorFigOff}{black!35}
    \colorlet{ColorFigOut}{ACMOrange}
    \colorlet{ColorFigConst}{ACMGreen!85!black}
    \begin{subfigure}[b]{.3\textwidth}
\begin{lstlisting}
#@CFG_EDGE: entry -> b1, b2
#@CFG_EDGE: b1, b2 -> exit
#@CONST_TB: 500:1, 0:2
#@INOUT_EX: codoku(10,20) == 1050
#@EXE_PATH: entry -> b1 -> exit
def codoku(a: int, b: int):
  #@CFG_BLOCK: entry
  c: int = -§\lstcodokucell{CONST}§
  if (§\lstcodokucell{ID}§ §\lstcodokucell{OP}§ §\lstcodokucell{ID}§) == -10:
    #@CFG_BLOCK: b1
    §\lstcodokucell{ID}§ += §\lstcodokucell{CONST}§
  else:
    #@CFG_BLOCK: b2
    §\lstcodokucell{ID}§ += §\lstcodokucell{CONST}§
  #@CFG_BLOCK: exit
  return a + 2 * b + c
\end{lstlisting}
    \caption{Puzzle}
    \end{subfigure}
    \hfill
    \begin{subfigure}[b]{.36\textwidth}
    \centering
    \hspace*{-9.8pt}%
    \begin{tikzpicture}[
        >={Stealth[length=3.2pt,width=2.6pt]},
        code/.style={font=\ttfamily\scriptsize, inner sep=0pt, text height=5pt, text depth=1.5pt},
        cell/.style={code, fill=ColorLstHighlight, inner xsep=1pt, inner ysep=.6pt, sharp corners},
        dom/.style={font=\ttfamily\tiny, text=black!60, inner sep=0pt, text height=3.6pt, text depth=1pt},
        stmt/.style={matrix of nodes, ampersand replacement=\&, column sep=1.2pt, row sep=.6pt,
                     inner sep=2.2pt, rounded corners=1.5pt, nodes={anchor=base}},
        io/.style={trapezium, trapezium left angle=72, trapezium right angle=108, trapezium stretches body,
                   font=\ttfamily\scriptsize, inner sep=1.5pt, text height=5pt, text depth=1.5pt},
        pathnode/.style={draw=ColorFigPath, line width=.6pt},
        offnode/.style={draw=ColorFigOff, text=ColorFigOff, line width=.5pt},
        offcell/.style={cell, fill=ColorLstHighlight!45, text=ColorFigOff},
        offdom/.style={dom, text=ColorFigOff},
        flow/.style={->, draw=ColorFigPath, line width=1.1pt},
        offflow/.style={draw=ColorFigOff, line width=.5pt},
        outlink/.style={->, draw=ColorFigOut, densely dashed, line width=.7pt},
        tblink/.style={draw=ColorFigConst, densely dotted, line width=.8pt},
        lbl/.style={font=\sffamily\tiny, inner sep=1.2pt},
        blk/.style={font=\sffamily\tiny, text=black!55, inner sep=.8pt},
        blkbg/.style={fill=black!6, rounded corners=2pt, inner sep=1.8pt},
    ]
    \node[io, pathnode] (in) at (0,0) {a=10, b=20};
    \matrix (s1) [stmt, pathnode, column sep=0pt] at (0,-.9) {
        |[code]| c = -\& |[cell]| CONST \\
                     \& |[dom]| \char123 500,0\char125 \\
    };
    \matrix (cond) [stmt, column sep=2.4pt] at (0,-1.9) {
        |[code]| ( \& |[cell]| ID \& |[cell]| OP \& |[cell]| ID \& |[code]| ) == -10 \\
        \& |[dom]| \char123 a,b,c\char125 \& |[dom]| \char123 +,-,*,{\rmfamily\ldots}\char125 \& \& |[dom]| {\sffamily\itshape e.g.,} a + c \\
    };
    \coordinate (condL) at ([xshift=-6pt]cond.west);
    \coordinate (condR) at ([xshift=6pt]cond.east);
    \draw[pathnode] (cond.north west) -- (cond.north east) -- (condR) -- (cond.south east)
        -- (cond.south west) -- (condL) -- cycle;
    \matrix (b1) [stmt, pathnode] at (-1.12,-3.0) {
        |[cell]| ID \& |[code]| += \& |[cell]| CONST \\
        |[dom]| \char123 a,b,c\char125 \& \& |[dom]| \char123 500,0\char125 \\
    };
    \matrix (b2) [stmt, offnode] at (1.12,-3.0) {
        |[offcell]| ID \& |[code]| += \& |[offcell]| CONST \\
        |[offdom]| \char123 a,b,c\char125 \& \& |[offdom]| \char123 500,0\char125 \\
    };
    \node[code, pathnode, rounded rectangle, inner xsep=3pt, inner ysep=2.5pt] (ret) at (0,-4.27) {return a+2*b+c};
    \node[io, draw=ColorFigOut] (o) at (1.9,-4.27) {1050};
    \begin{scope}[on background layer]
        \node[blkbg, fit=(s1)(cond)(condL)(condR)] (Bentry) {};
        \node[blkbg, fit=(b1)] (Bb1) {};
        \node[blkbg, fill=black!3, fit=(b2)] (Bb2) {};
        \node[blkbg, fit=(ret)] (Bexit) {};
    \end{scope}
    \node[blk, anchor=south west] at (Bentry.north west) {entry};
    \node[blk, anchor=south west] at (Bb1.north west) {b1};
    \node[blk, anchor=south east, text=ColorFigOff] at (Bb2.north east) {b2};
    \node[blk, anchor=south west] at (Bexit.north west) {exit};
    \coordinate (join) at ($(Bexit.north)+(0,.3)$);
    \draw[offflow] (cond.south -| b2.north) -- (b2.north) node[lbl, midway, right, text=ColorFigOff] {F};
    \draw[offflow] (b2.south) |- (join);
    \draw[flow] (in) -- (s1);
    \draw[flow] (s1) -- (cond.north);
    \draw[flow] (cond.south -| b1.north) -- (b1.north) node[lbl, midway, left, text=ColorFigPath] {T};
    \draw[flow] (b1.south) |- (join) -- (ret.north);
    \draw[->, draw=ColorFigOut, line width=.7pt] (ret) -- (o);
    \coordinate (busL) at ($(Bb1.west)+(-.2,0)$);
    \draw[outlink] (s1.west) -| (busL) |- (ret.west);
    \draw[outlink, -] (b1.west) -- (b1.west -| busL);
    \coordinate (busR) at ($(Bb2.east)+(.2,0)$);
    \node[blk, draw=ColorFigConst, text=black, rounded corners=1pt, align=center, font=\ttfamily\tiny, inner sep=1.5pt,
          anchor=north east] (tb) at ([xshift=.04cm]busR |- in.north) {CONST\_TB\\500:1, 0:2};
    \coordinate (busB) at ($(b2.south)!.5!(b2.south |- join)$);
    \draw[tblink] (tb.south -| busR) -- (busR |- busB);
    \draw[tblink] (s1-1-2.east) -- (s1-1-2.east -| busR);
    \draw[tblink] (b2-1-3.east) -- (b2-1-3.east -| busR);
    \draw[tblink] (b1-1-3 |- b1.south) |- (busR |- busB);
    \path (Bexit.south) ++(0,-5.2pt);
\end{tikzpicture}
    \caption{Core Structure}
    \end{subfigure}
    \hfill
    \begin{subfigure}[b]{.3\textwidth}
\begin{lstlisting}
#@CFG_EDGE: entry -> b1, b2
#@CFG_EDGE: b1, b2 -> exit
#@CONST_TB: 500:1, 0:2
#@INOUT_EX: codoku(10,20) == 1050
#@EXE_PATH: entry -> b1 -> exit
def codoku(a: int, b: int):
  #@CFG_BLOCK: entry
  c: int = -§\lstcodokucellsol{0}§
  if (§\lstcodokucellsol{c}§ §\lstcodokucellsol{-}§ §\lstcodokucellsol{a}§) == -10: # §\color{ColorCodeComment}\smash{\underline{or (a - b)}}§
    #@CFG_BLOCK: b1
    §\lstcodokucellsol{b}§ += §\lstcodokucellsol{500}§
  else:
    #@CFG_BLOCK: b2
    §\lstcodokucellsol{a}§ += §\lstcodokucellsol{0}§           # §\color{ColorCodeComment}\smash{\underline{or b, or c}}§
  #@CFG_BLOCK: exit
  return a + 2 * b + c
\end{lstlisting}
    \caption{Six Solutions}
    \end{subfigure}
    \caption{
      \textbf{Valid solutions are sparse}.
      (a) A toy puzzle with eight typed cells.
      (b) Its core structure, with one node per statement.
      On input \code{codoku(10,20)}, execution must follow \code{entry -> b1 -> exit} (blue), so \code{b2} never runs (grey).
      With three choices per \codokucell{ID}, two per \codokucell{CONST}, and 25 for \codokucell{OP}, there are $3^4\times2^3\times25 = 16{,}200$ candidate fillings.
      Global constraints couple distant cells:
      the return value must be \code{1050}, which ties the \code{return} to \code{c = -}\codokucell{CONST} and to the update in \code{b1} (orange),
      and the three \codokucell{CONST} cells share one constant table (green).
      (c) The six valid fillings: the branch condition admits two choices and the \codokucell{ID} in the unexecuted \code{b2} admits three; all other cells are forced.
    }
    \label{fig:codoku-example}
\end{figure*}

\Cref{fig:codoku-example} presents a toy codoku puzzle and its six solutions (see \Cref{sec:app_case} for a full-scale puzzle).
A solution must fill every \codokucell{ID}, \codokucell{CONST}, and \codokucell{OP} cell such that the completed program
\begin{enumerate}
    \item is accepted by the Python compiler;
    \item matches the control-flow graph (CFG) declared by the \code{\#@CFG\_EDGE} and \code{\#@CFG\_BLOCK} annotations;
    \item uses only constants listed in \code{\#@CONST\_TB}, each as often as specified (\code{500} once and \code{0} twice); and
    \item on input \code{codoku(10,20)}, returns \code{1050} (\code{\#@INOUT\_EX}) along the execution path (EP) in \code{\#@EXE\_PATH}.
\end{enumerate}
The first three constraints are static, whereas the fourth constrains the program's dynamic behavior.

\smalltitle{Reasoning by necessity}
Codokus make program reasoning unavoidable, even for agents equipped with tools.
First, a codoku puzzle has no behavior to observe:
being a partial program, it neither compiles nor runs, so execution, instrumentation, and debugging cannot reveal a valid filling.
Second, its search space grows exponentially with the number of cells, whereas valid solutions are sparse,
which renders random guessing and exhaustive enumeration impractical, if not infeasible;
even the toy puzzle in \Cref{fig:codoku-example} admits 16,200 fillings, only six of which are valid.
To fill a cell, a solver therefore \emph{must} characterize the program states that can reach the corresponding program point,
and then propagate the consequences of its choice forward along the prescribed execution path and backward from the required return value.
Because a locally plausible choice may surface as a violation only several statements or loop iterations later, cells cannot be resolved independently.

\smalltitle{Renewable by construction}
To generate codokus, we build on \emph{semantic reification}~\citep{reify}, a recent technique from the programming-languages community that synthesizes programs from scratch to satisfy prescribed semantic constraints.
Given a CFG, an input, and an EP, our synthesizer constructs a terminating program $P^\star$ that follows the path on the input.
Masking $P^\star$ with typed cells yields a puzzle, and $P^\star$ itself serves as a \emph{witness} that the puzzle has at least one valid solution.
Solutions, however, are verified against the global constraints rather than against the witness:
our solution checker parses a submitted filling, checks the constraints one by one, and accepts the filling if it satisfies all of them.
For the puzzle in \Cref{fig:codoku-example}, the checker thus accepts all six solutions.

Because puzzles are synthesized from scratch, they are independent of existing programs and renewable by construction, which addresses the supply problem:
fresh instances can always be drawn after a model is trained, reducing the risk that the exact programs and solutions have been seen during training or can be retrieved from the web.
The synthesizer exposes parameters that control puzzle complexity, including program size and the number of loop iterations.
Based on these parameters, we define three generation profiles:
a \code{small} codoku encodes a small straight-line function with a short execution path,
a \code{medium} codoku a function with shallow loops and a moderately long execution path,
and a \code{large} codoku a large branching function with nested loops.
Rather than assuming that these profiles directly determine difficulty, we measure how they affect agent performance.

\smalltitle{Evaluating frontier agents}
Codokus are designed for both bare models and coding agents, and we do not restrict how an agent solves them:
an agent may use any available tool to execute candidate fillings, invoke the solution checker, inspect failures, and implement its own search procedure in any programming language.
We therefore measure whether an agent finds a valid solution within a fixed resource budget, regardless of the method it uses.
We evaluate five frontier models with the Pi coding agent~\citep{piagent} on 300 codoku puzzles, 100 from each generation profile.
Codoku puzzles challenge coding agents:
even on small puzzles, solve rates range from 39\% to 77\%, and the highest solve rate on large puzzles is 54\%.
Every model solves fewer large puzzles than small ones, and resource use grows with profile scale.
Tool-use trajectories further show that agents combine multiple strategies to infer local program states to solve puzzles.
These results establish our contribution:
\emph{Codoku, a renewable testbed for program reasoning that can keep pace with rapidly improving coding agents}.

\section{Codoku Construction}\label{sec:codoku}

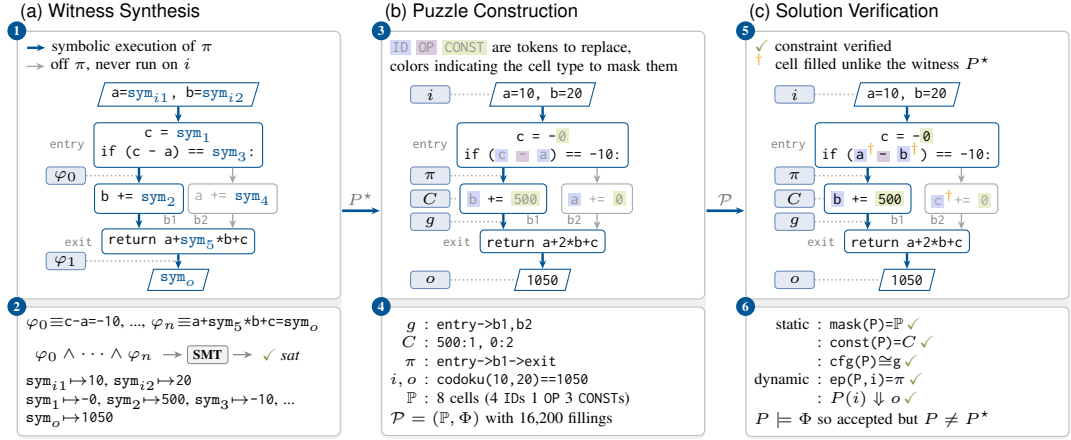
\begin{figure*}[t]
    \centering
    \colorlet{ColorFigPath}{ACMDarkBlue}
    \colorlet{ColorFigOff}{black!35}
    \colorlet{ColorFigOut}{ACMOrange}
    \colorlet{ColorFigOk}{ACMGreen!70!black}
    \colorlet{ColorCellID}{ColorLstHighlight}
    \colorlet{ColorCellOP}{ACMPurple!18}
    \colorlet{ColorCellCO}{ACMGreen!22}
    \DeclareRobustCommand{\figstep}[1]{\tikz[baseline=(ch.base)]{\node[shape=circle, fill=ColorFigPath,
        text=white, inner sep=.75pt, font=\fontsize{5}{5}\selectfont\bfseries] (ch) {#1};}}
    {\setlength{\fboxsep}{.9pt}%
    \hyphenpenalty=10000 \exhyphenpenalty=10000 \tolerance=9999 %
    \newcommand{\figtt}[1]{{\ttfamily\tiny#1}}%
    \newcommand{\cID}[1]{\colorbox{ColorCellID}{\ttfamily\tiny#1}}%
    \newcommand{\cOP}[1]{\colorbox{ColorCellOP}{\ttfamily\tiny#1}}%
    \newcommand{\cCO}[1]{\colorbox{ColorCellCO}{\ttfamily\tiny#1}}%
    \newcommand{\mID}[1]{\colorbox{ColorCellID}{\ttfamily\tiny\textcolor{black!42}{#1}}}%
    \newcommand{\mOP}[1]{\colorbox{ColorCellOP}{\ttfamily\tiny\textcolor{black!42}{#1}}}%
    \newcommand{\mCO}[1]{\colorbox{ColorCellCO}{\ttfamily\tiny\textcolor{black!42}{#1}}}%
    \newcommand{\symin}[1]{{\color{ColorFigPath}$\mathtt{sym}_{i#1}$}}%
    \newcommand{\symv}[1]{{\color{ColorFigPath}$\mathtt{sym}_{#1}$}}%
    \newcommand{\symo}{{\color{ColorFigPath}$\mathtt{sym}_{o}$}}%
    \newcommand{\psymin}[1]{$\mathtt{sym}_{i#1}$}%
    \newcommand{\psymv}[1]{$\mathtt{sym}_{#1}$}%
    \newcommand{\psymo}{$\mathtt{sym}_{o}$}%
    \newcommand{\figok}{{\color{ColorFigOk}$\checkmark$}}%
    \newcommand{\figdiff}{{\color{ColorFigOut}$^{\dagger}$}}%
    \scalebox{1.0}[1]{%
    \begin{tikzpicture}[
        font=\tiny,
        >={Stealth[length=3.2pt,width=2.6pt]},
        panel/.style={fill=black!6, rounded corners=3pt, inner sep=3.5pt},
        phead/.style={font=\sffamily\scriptsize, text=black, inner sep=1.6pt},
        frame/.style={draw=black!25, fill=white, rounded corners=2pt, inner sep=3pt},
        tbox/.style={align=left, anchor=north west, text width=4.046cm, inner sep=0pt},
        num/.style={circle, fill=ColorFigPath, text=white, inner sep=0pt,
                    minimum size=2.5mm, font=\fontsize{5}{5}\selectfont\bfseries},
        blk/.style={draw=ColorFigPath, fill=white, rounded corners=1.5pt, align=center,
                    font=\ttfamily\tiny, inner xsep=2.5pt, inner ysep=1.8pt},
        blkoff/.style={blk, draw=ColorFigOff, text=ColorFigOff},
        io/.style={draw=ColorFigPath, fill=white, trapezium, trapezium left angle=72,
                   trapezium right angle=108, trapezium stretches body, font=\ttfamily\tiny,
                   inner xsep=2.5pt, inner ysep=1.2pt},
        rIN/.style={io, minimum height=.31cm},
        rEN/.style={blk, minimum height=.59cm},
        rBB/.style={blk, minimum height=.39cm},
        rBBoff/.style={blkoff, minimum height=.39cm},
        rEX/.style={blk, minimum height=.30cm},
        rOUT/.style={io, minimum height=.28cm},
        ed/.style={->, draw=ColorFigPath, line width=.9pt},
        edoff/.style={->, draw=ColorFigOff, line width=.5pt},
        stag/.style={draw=ColorFigPath!55, fill=ColorFigPath!8, rounded corners=1.2pt,
                     font=\tiny, inner xsep=1.6pt, inner ysep=1pt, anchor=east,
                     minimum width=.52cm, minimum height=.23cm, align=center},
        tlead/.style={draw=black!35, densely dotted, line width=.55pt},
        smtchip/.style={draw=black!45, fill=black!5, rounded corners=1pt,
                    font=\fontsize{5}{5}\selectfont\bfseries,
                    inner xsep=2.2pt, inner ysep=1.6pt},
        flow/.style={->, draw=ColorFigPath, line width=1pt},
        thinar/.style={->, draw=black!45, line width=.5pt},
        cmt/.style={font=\fontsize{5.6}{6.4}\selectfont\itshape, text=black!55, inner sep=1pt},
        elbl/.style={font=\fontsize{5.2}{5.2}\selectfont\ttfamily, text=black!50, inner sep=1.2pt},
    ]

    \coordinate (aL) at (0.137,-0.13);
    \coordinate (aR) at (4.183,-0.13);
    \node[tbox] (a1t) at (0.137,-0.13)
        {\tikz[baseline=-.4ex]{\draw[ed] (0,0) -- (.24,0);} symbolic execution of $\pi$\\
         \tikz[baseline=-.4ex]{\draw[edoff] (0,0) -- (.24,0);} off $\pi$, never run on $i$};
    \node[rIN]    (Ain)  at (2.160,-0.86) {a=\symin{1}, b=\symin{2}};
    \node[rEN]    (Aen)  at (2.160,-1.53) {c = \symv{1}\\if (c - a) == \symv{3}:};
    \node[rBB]    (Ab1)  at (1.640,-2.24) {b += \symv{2}};
    \node[rBBoff] (Ab2)  at (2.880,-2.24) {a += \symv{4}};
    \node[rEX]    (Aex)  at (2.160,-2.805) {return a+\symv{5}*b+c};
    \node[rOUT]   (Aout) at (2.160,-3.315) {\symo};
    \draw[ed]    (Ain) -- (Aen);
    \draw[ed]    (Aen.south -| Ab1) -- (Ab1.north);
    \draw[edoff] (Aen.south -| Ab2) -- (Ab2.north);
    \draw[ed]    (Ab1.south) -- (Aex.north -| Ab1);
    \draw[edoff] (Ab2.south) -- (Aex.north -| Ab2);
    \draw[ed]    (Aex) -- (Aout);
    \node[elbl, anchor=east] (Aenl) at ([xshift=-2pt]Aen.west) {entry};
    \node[elbl, anchor=east] (Aexl) at ([xshift=-2pt]Aex.west) {exit};
    \node[elbl, anchor=north east] (Ab1l) at ([xshift=-1pt,yshift=1pt]Ab1.south east) {b1};
    \node[elbl, anchor=north west] (Ab2l) at ([xshift=1pt,yshift=1pt]Ab2.south west) {b2};
    \node[stag] (Ap0) at (0.940,-1.935) {$\varphi_0$};
    \node[stag] (Ap1) at (0.940,-3.065) {$\varphi_1$};
    \draw[tlead] (Ap0.east) -- (Ab1.north |- Ap0);
    \draw[tlead] (Ap1.east) -- (Aout.north |- Ap1);

    \node[tbox] (a2t) at (0.137,-3.78) {%
        $\varphi_0$$\equiv$\figtt{c-a=-10},
        ...,
        $\varphi_n$$\equiv$\figtt{a+\psymv{5}*b+c=}\psymo        
    };
    \node[anchor=north west] (Aq) at ([yshift=-3pt]a2t.south west) {$\varphi_0\wedge\cdots\wedge\varphi_n$};
    \node[smtchip, anchor=west] (Asmt) at ([xshift=9pt]Aq.east) {SMT};
    \draw[thinar] (Aq.east) -- ([xshift=-.06cm]Asmt.west);
    \node[anchor=west] (Asat) at ([xshift=9pt]Asmt.east) {\figok\ \emph{sat}};
    \draw[thinar] ([xshift=.06cm]Asmt.east) -- (Asat.west);
    \coordinate (arow) at ([yshift=-4pt]Asmt.south);
    \node[tbox] (a3t) at (a2t.west |- arow)
        {\psymin{1}\figtt{$\mapsto$10}, \psymin{2}\figtt{$\mapsto$20}\\
         \psymv{1}\figtt{$\mapsto$-0}, \psymv{2}\figtt{$\mapsto$500}, \psymv{3}\figtt{$\mapsto$-10}, ...\\
         \psymo\figtt{$\mapsto$1050}};

    \coordinate (bL) at (4.957,-0.13);
    \coordinate (bR) at (9.003,-0.13);
    \node[tbox] (b1t) at (4.957,-0.13)
        {\mID{ID} \mOP{OP} \mCO{CONST} are tokens to replace,\\
         colors indicating the cell type to mask them};
    \node[rIN] (Bin) at (6.980,-0.86) {a=10, b=20};
    \node[rEN]    (Ben)  at (6.980,-1.53) {c = -\mCO{0}\\if (\mID{c} \mOP{-} \mID{a}) == -10:};
    \node[rBB]    (Bb1)  at (6.460,-2.24) {\mID{b} += \mCO{500}};
    \node[rBBoff] (Bb2)  at (7.700,-2.24) {\mID{a} += \mCO{0}};
    \node[rEX]    (Bex)  at (6.980,-2.805) {return a+2*b+c};
    \node[rOUT] (Bout) at (6.980,-3.315) {1050};
    \draw[ed]    (Bin) -- (Ben);
    \draw[ed]    (Ben.south -| Bb1) -- (Bb1.north);
    \draw[edoff] (Ben.south -| Bb2) -- (Bb2.north);
    \draw[ed]    (Bb1.south) -- (Bex.north -| Bb1);
    \draw[edoff] (Bb2.south) -- (Bex.north -| Bb2);
    \draw[ed] (Bex) -- (Bout);
    \node[elbl, anchor=east] (Benl) at ([xshift=-2pt]Ben.west) {entry};
    \node[elbl, anchor=east] (Bexl) at ([xshift=-2pt]Bex.west) {exit};
    \node[elbl, anchor=north east] (Bb1l) at ([xshift=-1pt,yshift=1pt]Bb1.south east) {b1};
    \node[elbl, anchor=north west] (Bb2l) at ([xshift=1pt,yshift=1pt]Bb2.south west) {b2};
    \node[stag] (Bi)  at (5.76,-0.86)  {$i$};
    \node[stag] (Bpi) at (5.76,-1.935) {$\pi$};
    \node[stag] (Bc)  at (5.76,-2.24)  {$C$};
    \node[stag] (Bg)  at (5.76,-2.545) {$g$};
    \node[stag] (Bo)  at (5.76,-3.315) {$o$};
    \draw[tlead]   (Bi.east)  -- (Bin.west);
    \draw[tlead]   (Bpi.east) -- (Bb1.north |- Bpi);
    \draw[tlead] (Bc.east)  -- (Bb1.west);
    \draw[tlead]                        (Bg.east)  -- (Bb1.south |- Bg);
    \draw[tlead]   (Bo.east)  -- (Bout.west);

    \node[tbox] (b2t) at (4.957,-3.78) {%
      \begin{tabular}{@{}r@{\enspace:\enspace}l@{}}
        $g$                              & \figtt{entry->b1,b2} \\
        $C$                              & \figtt{500:1, 0:2} \\
        $\pi$                            & \figtt{entry->b1->exit} \\
        $i,o$                            & \figtt{codoku(10,20)==1050} \\
        $\mathbb{P}$                     & 8 cells (4 \figtt{ID}s 1 \figtt{OP} 3 \figtt{CONST}s)
      \end{tabular}\\
      $\mathcal{P}=(\mathbb{P},\Phi)$ with 16,200 fillings
    };

    \coordinate (cL) at (9.777,-0.13);
    \coordinate (cR) at (13.823,-0.13);
    \node[tbox] (c1t) at (9.777,-0.13)
        {\figok\ constraint verified\\
         \figdiff\ \ cell filled unlike the witness $P^\star$};
    \node[rIN] (Cin) at (11.800,-0.86) {a=10, b=20};
    \node[rEN]    (Cen)  at (11.800,-1.53) {c = -\cCO{0}\\if (\cID{a}\figdiff \cOP{-} \cID{b}\figdiff) == -10:};
    \node[rBB]    (Cb1)  at (11.280,-2.24) {\cID{b} += \cCO{500}};
    \node[rBBoff] (Cb2)  at (12.520,-2.24) {\cID{c}\figdiff += \cCO{0}};
    \node[rEX]    (Cex)  at (11.800,-2.805) {return a+2*b+c};
    \node[rOUT] (Cout) at (11.800,-3.315) {1050};
    \draw[ed]    (Cin) -- (Cen);
    \draw[ed]    (Cen.south -| Cb1) -- (Cb1.north);
    \draw[edoff] (Cen.south -| Cb2) -- (Cb2.north);
    \draw[ed]    (Cb1.south) -- (Cex.north -| Cb1);
    \draw[edoff] (Cb2.south) -- (Cex.north -| Cb2);
    \draw[ed] (Cex) -- (Cout);
    \node[elbl, anchor=east] (Cenl) at ([xshift=-2pt]Cen.west) {entry};
    \node[elbl, anchor=east] (Cexl) at ([xshift=-2pt]Cex.west) {exit};
    \node[elbl, anchor=north east] (Cb1l) at ([xshift=-1pt,yshift=1pt]Cb1.south east) {b1};
    \node[elbl, anchor=north west] (Cb2l) at ([xshift=1pt,yshift=1pt]Cb2.south west) {b2};
    \node[stag] (Ci)  at (10.580,-0.86)  {$i$};
    \node[stag] (Cpi) at (10.580,-1.935) {$\pi$};
    \node[stag] (Cc)  at (10.580,-2.24)  {$C$};
    \node[stag] (Cg)  at (10.580,-2.545) {$g$};
    \node[stag] (Co)  at (10.580,-3.315) {$o$};
    \draw[tlead]   (Ci.east)  -- (Cin.west);
    \draw[tlead]   (Cpi.east) -- (Cb1.north |- Cpi);
    \draw[tlead] (Cc.east)  -- (Cb1.west);
    \draw[tlead]                        (Cg.east)  -- (Cb1.south |- Cg);
    \draw[tlead]   (Co.east)  -- (Cout.west);

    \node[tbox] (c2t) at (9.777,-3.78) {%
      \begin{tabular}{@{}r@{\enspace:\enspace}l@{}}
        static             & \figtt{mask(P)=$\mathbb{P}$}\,\figok \\
                           & \figtt{const(P)=$C$}\,\figok \\
                           & \figtt{cfg(P)$\cong$g}\,\figok \\
        dynamic            & \figtt{ep(P,i)=$\pi$}\,\figok \\
                           & $P(i) \Downarrow o$\,\figok\\
      \end{tabular}\\
      $P\models\Phi$ so accepted but $P \neq P^\star$
    };

    \coordinate (padA) at (2.160,-5.30);
    \coordinate (padB) at (6.980,-5.30);
    \coordinate (padC) at (11.800,-5.30);

    \begin{scope}[on background layer]
        \node[panel, fit=(a1t)(aL)(aR)(Ain)(Aen)(Aex)(Ab2l)(Aout)(Aenl)(Ap0)(a2t)(a3t)(Asat)(padA)] (PA) {};
        \node[panel, fit=(b1t)(bL)(bR)(Bin)(Ben)(Bex)(Bb2l)(Bout)(Benl)(Bi)(b2t)(padB)] (PB) {};
        \node[panel, fit=(c1t)(cL)(cR)(Cin)(Cen)(Cex)(Cb2l)(Cout)(Cenl)(Ci)(c2t)(padC)] (PC) {};
        \node[frame, fit=(a1t)(aL)(aR)(Ain)(Aen)(Aex)(Ab2l)(Aout)(Aenl)(Ap0)(Ap1)] (a1) {};
        \node[frame, fit=(a2t)(a3t)(Aq)(Asmt)(Asat)] (a2) {};
        \node[frame, fit=(b1t)(bL)(bR)(Bin)(Ben)(Bex)(Bb2l)(Bout)(Benl)(Bi)(Bo)] (b1) {};
        \node[frame, fit=(b2t)] (b2) {};
        \node[frame, fit=(c1t)(cL)(cR)(Cin)(Cen)(Cex)(Cb2l)(Cout)(Cenl)(Ci)(Co)] (c1) {};
        \node[frame, fit=(c2t)] (c2) {};
    \end{scope}
    \node[phead, anchor=south west] at ([yshift=1.6pt]PA.north west) {(a) Witness Synthesis};
    \node[phead, anchor=south west] at ([yshift=1.6pt]PB.north west) {(b) Puzzle Construction};
    \node[phead, anchor=south west] at ([yshift=1.6pt]PC.north west) {(c) Solution Verification};

    \foreach \i/\n in {a1/1,a2/2,b1/3,b2/4,c1/5,c2/6}
        \node[num] at (\i.north west) {\n};

    \coordinate (mid) at (0,-2.40);
    \draw[flow] (PA.east |- mid) -- node[cmt, above, inner sep=1.5pt] {$P^\star$} (PB.west |- mid);
    \draw[flow] (PB.east |- mid) -- node[cmt, above, inner sep=1.5pt] {$\mathcal{P}$} (PC.west |- mid);

    \end{tikzpicture}}}
    \caption{
      \textbf{Building a codoku puzzle and checking a solution} (for \Cref{fig:codoku-example}).
      \textbf{(a)} Semantic reification populates a required CFG with symbolic inputs, intermediates, and output~\figstep{1},
      encodes the conditions imposed by the path $\pi$, and solves them in one SMT query~\figstep{2},
      returning a witness solution $P^\star$.
      \textbf{(b)} Masking replaces shaded tokens with typed cells~\figstep{3}.
      The witness supplies the constraints: its graph gives $g$, its constants the constant table $C$, and its run on $i$ the path $\pi$ and output $o$~\figstep{4}.
      \textbf{(c)} A candidate is checked in the same places~\figstep{5}: first statically (cells, structure, constants), then dynamically by running it on $i$~\figstep{6}.
      Checking against $\Phi$, not $P^\star$, accepts valid fillings that differ from the witness.
    }
    \label{fig:codoku-pipeline}
\end{figure*}

Codoku formulates program reasoning as constraint satisfaction over a masked program.
Constructing a puzzle therefore requires producing a partial program, together with static and dynamic semantic constraints, that jointly meet four requirements:
\begin{enumerate}
\item every puzzle admits at least one witness solution by construction (\emph{solvable});
\item its solution space is large while its valid solutions are few (\emph{challenging});
\item a candidate solution can be verified objectively in polynomial time (\emph{verifiable});
\item fresh puzzles can be drawn on demand (\emph{renewable}).
\end{enumerate}
\Cref{fig:codoku-pipeline} illustrates the pipeline on the puzzle of \Cref{fig:codoku-example}:
witness synthesis (\Cref{ssec:witness_synthesis}), puzzle construction (\Cref{ssec:puzzle_construction}), and solution verification (\Cref{ssec:solution_verification}).

\subsection{Witness Synthesis}\label{ssec:witness_synthesis}

Most program generators are syntax-guided:
they emit syntactically correct programs but offer no guarantees about runtime behavior on a specific input,
so their output may raise unhandled exceptions, enter infinite loops, or consist largely of dead code.
For example, Csmith~\citep{csmith} guarantees semantic correctness, yet does not offer precise control over semantic properties such as loop behavior and execution length.
Lacking such control, the resulting codoku puzzles would be too easy or too hard.
We therefore build on \emph{semantic reification}~\citep{reify}, which
synthesizes programs that satisfy prescribed static and dynamic semantic specifications.

\smalltitle{Semantic reification}
Let $g = ( V, E )$ be a control-flow graph (CFG), where
$V$ denotes the set of basic blocks, among them a unique $\code{entry}$ and a unique $\code{exit}$ block,
and $E \subseteq V \times V$ the set of directed control-flow edges.
An execution path (EP) through $g$ is a finite walk $\pi = [b_1, b_2, \dots, b_{|\pi|}]$ from the entry to the exit, \ie, $b_1 = \code{entry}$ and $b_{|\pi|} = \code{exit}$.
From a given CFG $g$ and EP $\pi$, semantic reification synthesizes a program $P^\star$ together with an input $i$ and the corresponding output $o$:
$$
\begin{aligned}
\code{reify}(g,\pi) &\leadsto (P^\star,i,o) \\
&\textit{s.t.} \\
\code{cfg}(P^\star) \cong g,
P^\star(i) &\Downarrow o,
\code{ep}(P^\star,i) = \pi.
\end{aligned}
$$
That is, $P^\star$'s CFG is isomorphic to $g$,
executing $P^\star$ on $i$ terminates deterministically with output $o$, and
the sequence of basic blocks traversed by that execution is exactly $\pi$.

\smalltitle{Witness synthesis}
We instantiate semantic reification in the following four steps:
\begin{enumerate}
    \item \emph{CFG and EP creation}:
    We construct a connected CFG $g$ by adding forward branches and loop back edges,
    then walk it to obtain the EP $\pi$.

    \item \emph{Statement population}:
    We populate every basic block with symbolic Python statements,
    seeding the blocks along $\pi$ with statements over \emph{symbolic} variables and filling those outside $\pi$ with concrete statements.
    Off-path code is never executed on the input $i$, which Step 4 determines.

    \item \emph{Symbolic execution}:
    We symbolically execute the populated program along $\pi$, encoding the path, definedness, and computation conditions required to follow $\pi$ as bit-vector constraints.

    \item \emph{Program concretization}:
    We solve these constraints with an SMT solver, obtaining concrete values for the input $i$, the output $o$, and every symbolic variable.
    Substituting each symbolic variable with its solved value yields the final program $P^\star$.
\end{enumerate}
$P^\star$ is the witness solution from which we construct a codoku puzzle,
while the CFG $g$, EP $\pi$, input $i$, and output $o$ are retained as constraints for solution verification.

\subsection{Puzzle Construction}\label{ssec:puzzle_construction}

Our puzzle synthesizer turns the witness program $P^\star$ into a codoku puzzle $\mathcal{P} = (\mathbb{P}, \Phi)$ in three steps:
it (1) replaces selected tokens with typed cells to form the puzzle body $\mathbb{P}$,
(2) collects the masked constants into a table, and
(3) records the global constraints $\Phi$ as annotations.

\smalltitle{Typed cells}
The synthesizer preserves $P^\star$'s function signature.
Within its body, it selects eligible statements and replaces their maskable tokens with typed cells, yielding $\mathbb{P}$:
\begin{itemize}
    \item \codokucell{ID}: A local variable or parameter identifier such as \code{v1}.
    \item \codokucell{FUNC}: A function name already available in the puzzle such as \code{sat\_add}.
    \item \codokucell{CONST}: A numeric constant, either an integer or a floating-point number such as \code{500}.
    \item \codokucell{OP}: A supported operator or conditional-expression keyword such as \code{+}.
    \item \codokucell{CTRL}: A control-flow keyword, either \code{break} or \code{continue}.
    \item \codokucell{LABEL}: A destination basic-block label such as \code{b1}.
\end{itemize}
Not every puzzle contains all six kinds.
A cell's type restricts its fillings:
\codokucell{ID} and \codokucell{FUNC} cells admit only variables and functions declared in the puzzle, respectively, and \codokucell{CONST} cells only table constants.

\smalltitle{Constant table}
The synthesizer collects the constants replaced by \codokucell{CONST} into a table
$$
C = \{(c_1, n_1), (c_2, n_2), \dots, (c_{|C|}, n_{|C|})\},
$$
where $c_j$ is a constant and $n_j$ its required number of occurrences.
Constants are distinguished by value and by type, so that the integer \code{1} and the floating-point \code{1.0} are distinct.
A valid solution must assign each $c_j$ to exactly $n_j$ constant cells and may use no constant outside the table.
Spending a constant in one cell thus reduces its remaining occurrences elsewhere, which prevents a solver from satisfying statements independently.
This budget, inspired by sudoku, rules out the arbitrary literals that would otherwise make individual cells easy to satisfy.

\smalltitle{Global constraints}
The synthesizer records global constraints $\Phi = (g, C, i, \pi, o)$ as annotations alongside the puzzle:
\code{\#@CFG\_\{BLOCK,EDGE\}} declare basic blocks and control-flow edges,
\code{\#@CONST\_TB} displays the constant table, and
\code{\#@INOUT\_EX} and \code{\#@EXE\_PATH} give the input-output example and the required EP.
Cell types in $\mathbb{P}$, such as \codokucell{ID}, further limit the admissible choices.
These constraints serve complementary roles:
the CFG constrains the program's static structure, while the EP and the output constrain its runtime behavior.
Together they define an explicit, constrained program-reasoning task, rather than an attempt to infer an unspecified program from input-output examples.

\subsection{Solution Verification}\label{ssec:solution_verification}
The witness $P^\star$ establishes that at least one valid filling exists, but a solver need not recover its particular choices.
The checker accordingly evaluates a candidate against the global constraints rather than against $P^\star$, and accepts every satisfying filling,
\ie, $\code{valid\_sols}(\mathcal{P}) = \{P \mid P \models \Phi \}$, where
$$
\begin{aligned}
P \models \Phi &{} \iff
     \bigl(\code{mask}(P) = \mathbb{P}\bigr) \\
&\quad\land\quad\bigl(\code{const}(P) = C\bigr) \\
&\quad\land\quad\bigl(\code{cfg}(P) \cong g\bigr) \\
&\quad\land\quad\bigl(P(i) \Downarrow o\bigr) \\
&\quad\land\quad\bigl(\code{ep}(P,i) = \pi\bigr).
\end{aligned}
$$

\smalltitle{Static constraints}
The checker first confirms that every cell is filled and that the candidate parses and compiles.
It then traverses the candidate's abstract syntax tree (AST) to perform four checks:
\begin{itemize}
    \item \emph{Cell and structure preservation}:
    Every cell must be filled with a token of its type, and all unmasked code must remain unchanged.
    The checker masks $P$ by the procedure that produced the typed cells of $P^\star$ (\Cref{ssec:puzzle_construction});
    the result must be identical to the puzzle at the AST level, comments ignored.
    Formatting changes are thus permitted, whereas altering an unmasked statement or inserting new code, such as replacing the puzzle body with a hardcoded return value, is not.
    \item \emph{Constant consistency}:
    Cell constants must match the constant table in value, type, and count.
    \item \emph{Declaration consistency}:
    Variables and functions used to fill cells must be declared in the puzzle.
    \item \emph{CFG consistency}:
    The CFG reconstructed from the candidate must match the declared CFG $g$.
\end{itemize}

\smalltitle{Dynamic constraints}
The checker executes the candidate on $i$ and records the sequence of basic blocks it visits.
The candidate must terminate with output $o$, and the recorded sequence must match the EP $\pi$ exactly, including repeated visits.
Output agreement alone is therefore insufficient:
a candidate that returns $o$ but skips a required block is invalid, as is one that follows $\pi$ but returns a different value.
A 5-second execution time limit further rejects candidates that do not finish in time.

\smalltitle{Checker feedback}
The checker stops at the first failed check and reports the violated constraint,
for instance a missing or unexpected CFG edge, a mismatch in the execution path, or an incorrect constant count.
Such feedback lets agents revise a candidate in response to a specific failure,
yet it neither discloses the witness's choices nor requires agents to reproduce them.

\subsection{Generation Profiles}\label{ssec:generation_profiles}
The complexity of a puzzle can be measured in several respects, of which we consider two:
search-space complexity and program complexity.
As in sudoku, the search-space complexity of a puzzle is the number of assignments admissible to its cells, that is, the size of the Cartesian product of the cells' domains;
it counts all permitted fillings, not only those that satisfy the puzzle constraints.
Program complexity metrics, widely used in software engineering and programming languages to estimate how difficult a program is to understand, test, maintain, or modify,
describe both static program structure and dynamic execution behavior, capturing properties that cell counts alone omit.
Following existing literature~\citep{progcomplmetrics} and the metrics plugins popular for
VS Code~\citep{vscodecodemetrics},
IntelliJ Platforms~\citep{intellijmetricsreloaded}, and
Eclipse~\citep{eclipsedepdigger},
we adopt for example
the numbers of statements and variables,
McCabe's cyclomatic complexity~\citep{mccabe},
the length of execution paths, and
the number of loop iterations,
which distinguish code size from the amount of execution on the specified input.
The synthesizer exposes both as parameters, from which we derive three profiles:
\begin{itemize}
    \item \textbf{Small}:
    Generate small straight-line functions with short execution.

    \item \textbf{Medium}:
    Generate larger functions with shallow loops and moderate execution.

    \item \textbf{Large}:
    Generate even larger branching functions with nested loops and long execution.
\end{itemize}

\Cref{ssec:rq1} reports each profile's empirical complexity.
A more complex puzzle is generally harder, but not always:
a puzzle with a large search space may be easy if local constraints eliminate most choices.
We therefore assess difficulty through agent success rates and solution costs (\Cref{ssec:rq2}).

\section{Evaluation}\label{sec:evaluation}

Because codokus are renewable by construction, our evaluation focuses on whether they are challenging
for frontier LLMs operating through a coding agent.
We address two research questions:
\begin{itemize}
    \item \textbf{RQ1: Codoku characterization}.
    Do the small, medium, and large profiles yield puzzles of increasing search space and program complexity?
    \item \textbf{RQ2: Challenge for coding agents}.
    How successfully do current coding agents solve codokus within a fixed resource budget, and how does their success vary across the generation profiles?
\end{itemize}

\subsection{Evaluation Setup}\label{ssec:evaluation_setup}

We generate 100 puzzles per profile, 300 in total.
Each puzzle comes with a witness solution, which is unavailable to agents.
We evaluate Claude Opus 5, GPT 5.6 Sol, GLM 5.3, Kimi K3, and DeepSeek V4.1 Flash, each with its thinking effort set to \code{high}.
All models run in the Pi coding agent~\citep{piagent} under an identical configuration.
The agent has access to the solution checker (\Cref{ssec:solution_verification}) and all standard tools,
and each agent is isolated in its own E2B micro VM with four CPU cores and 16~GB of memory.
Every model attempts all 300 puzzles (1,500 runs in total).

\smalltitle{Budget}
For each puzzle, we grant the agent at most one hour, \$15 in model usage, and 128 model requests.
A run ends when the agent submits a solution or exhausts any of these budgets.

\subsection{RQ1: Codoku Characterization}\label{ssec:rq1}

\begin{table*}[tb]
    \centering
    \footnotesize
    \caption{
        \textbf{Codokus challenge every evaluated agent, and larger profiles are more complex and generally harder}.
        The left table displays average puzzle properties per profile.
        The right table shows mutually exclusive run outcomes (\Cref{ssec:rq2}):
        a run is \emph{exhausted} when it reaches its time, cost, or request budget,
        and \emph{failed} when it otherwise ends without a valid solution.
    }
    \begin{subtable}[t]{.45\linewidth}
    \centering
    \caption{Puzzle Complexity}
    \label{tab:puzzle_complexity}
    \begin{tabular}{lrrr}
\toprule
\bf Metric            & \bf Small & \bf Medium & \bf Large \\
\cmidrule(lr){1-4}
\multicolumn{4}{c}{\textbf{\textit{\underline{search space}}}}
\vspace{3pt} \\
Typed cells           & 58.54  & 134.65 & 366.49 \\
Log10 space           & 71.99  & 182.26 & 540.03 \\
\cmidrule(lr){1-4}
\multicolumn{4}{c}{\textbf{\textit{\underline{static structure}}}}
\vspace{3pt} \\
Lines of code         & 38.85 &  71.94 & 133.46 \\
Halstead diff.        & 38.62 &  58.31 &  86.18 \\
CFG nodes             &  4.77 &   6.58 &   9.93 \\
CFG edges             &  5.47 &   8.91 &  14.60 \\
Cyclo. compl.         &  2.70 &   4.33 &   6.67 \\
Data dep. nodes       & 21.35 &  44.01 &  92.95 \\
Data dep. edges       & 16.74 &  46.29 & 111.66 \\
Data dep. degree      &  1.50 &   2.07 &   2.38 \\
\cmidrule(lr){1-4}
\multicolumn{4}{c}{\textbf{\textit{\underline{dynamic behavior}}}}
\vspace{3pt} \\
EP length             &  5.85 &   9.71 &  15.29 \\
Unique blocks         &  4.70 &   5.92 &   7.07 \\
Repeated blocks       &  1.15 &   3.79 &   8.22 \\
Loop iterations       &  1.12 &   2.52 &   4.09 \\
\cmidrule(lr){1-4}
\multicolumn{4}{c}{\textbf{\textit{\underline{input-output examples}}}}
\vspace{3pt} \\
Examples              &  3.71 &   6.30 &   9.00 \\
\bottomrule
    \end{tabular}
    \end{subtable}
    \hfill
    \begin{subtable}[t]{.52\linewidth}
    \centering
    \caption{Model Performance with the Pi Agent}
    \label{tab:agent_results}
    \begin{tabular}{llccc}
\toprule
\textbf{Model} &
\textbf{Profile} &
\textbf{Solved} &
\textbf{Exhausted} &
\textbf{Failed} \\
\cmidrule(lr){1-5}
\multirow{4}{*}{\makecell[l]{\bf Claude\\Opus 5}}
  & \textbf{\textit{\underline{Total}}}  & 63\% & 33\% & 4\% \\
  & \it $\cdot$~small                    & 77\% & 20\% & 3\% \\
  & \it $\cdot$~medium                   & 62\% & 33\% & 5\% \\
  & \it $\cdot$~large                    & 50\% & 45\% & 5\% \\
\cmidrule(lr){1-5}
\multirow{4}{*}{\makecell[l]{\bf GPT\\5.6 Sol}}
  & \textbf{\textit{\underline{Total}}}  & 58\% & 36\% &  6\% \\
  & \it $\cdot$~small                    & 67\% & 33\% &  0\% \\
  & \it $\cdot$~medium                   & 53\% & 43\% &  4\% \\
  & \it $\cdot$~large                    & 54\% & 33\% & 13\% \\
\cmidrule(lr){1-5}
\multirow{4}{*}{\makecell[l]{\bf GLM\\5.3}}
  & \textbf{\textit{\underline{Total}}}  & 28\% & 58\% & 14\% \\
  & \it $\cdot$~small                    & 50\% & 40\% & 10\% \\
  & \it $\cdot$~medium                   & 22\% & 53\% & 25\% \\
  & \it $\cdot$~large                    & 12\% & 80\% & 8\% \\
\cmidrule(lr){1-5}
\multirow{4}{*}{\makecell[l]{\bf Kimi\\K3}}
  & \textbf{\textit{\underline{Total}}}  & 29\% & 70\% & 1\% \\
  & \it $\cdot$~small                    & 48\% & 51\% & 1\% \\
  & \it $\cdot$~medium                   & 28\% & 72\% & 0\% \\
  & \it $\cdot$~large                    & 11\% & 89\% & 0\% \\
\cmidrule(lr){1-5}
\multirow{4}{*}{\makecell[l]{\bf DPSK\\V4.1\\Flash}}
  & \textbf{\textit{\underline{Total}}}  & 30\% & 64\% &  6\% \\
  & \it $\cdot$~small                    & 39\% & 61\% &  0\% \\
  & \it $\cdot$~medium                   & 29\% & 65\% &  6\% \\
  & \it $\cdot$~large                    & 22\% & 66\% & 12\% \\
\bottomrule
    \end{tabular}        
    \end{subtable}
\end{table*}

Every metric in \Cref{tab:puzzle_complexity} increases monotonically from the small to the medium to the large profile.
Profiles scale both dimensions of complexity (\Cref{ssec:generation_profiles}):
the search space and program complexity as reflected in static structure, dynamic behavior, and the number of input-output examples.

The profiles do not, however, merely lengthen the programs.
From the small to the large profile, code size grows 3.4$\times$ but the number of typed cells 6.3$\times$, so each line carries nearly twice as many cells;
the average data-dependency degree rises by more than 50\%, so each value takes part in more dependencies;
and repeated block visits, which make up 20\% of the execution path in small puzzles, exceed 53\% of it in large ones.
Larger puzzles therefore contain not only more decisions but also more relations among them.
Execution paths grow mainly through iteration, so each decision propagates through more loop iterations.
Larger profiles also supply more input-output examples, each both a further constraint on the filling and a further source of information (details in \Cref{sec:app_complexity}).

\smalltitle{Answer to RQ1}
The realized complexity follows the intended ordering:
the profiles scale both a solver's choices and the semantic dependencies that constrain them.
They do so without repository-scale programs: even large puzzles average 133 lines of code,
whereas even small puzzles span on the order of $10^{72}$ candidate fillings.
RQ2 asks whether such compact puzzles challenge current agents.

\subsection{RQ2: Challenge for Coding Agents}\label{ssec:rq2}

We classify each run into one of three mutually exclusive outcomes (\Cref{tab:agent_results}).
Additionally, a run is \emph{failed} if it ends without a valid solution for any other reason:
a safety refusal by the provider, an output-length error, an infrastructure failure, or the agent stopping without a valid completion.

\smalltitle{Compact puzzles pose substantial challenges}
No evaluated agent saturates codoku.
Claude Opus 5 reaches the highest overall solve rate, followed by GPT 5.6 Sol,
yet even Claude Opus 5 leaves almost a quarter of the small puzzles unsolved, although these average fewer than 40 lines of code.
No agent solves more than 54\% of the large ones.
Substantial headroom thus remains at every profile, even though the agents can execute code, validate candidates, and program their own search.

\smalltitle{A gap exist between proprietary and open-weight models}
The two proprietary models solve roughly twice as many puzzles as the three open-weight ones,
and the separation holds within every profile: both proprietary models outperform all three open-weight models on small, medium, and large puzzles.
Averaged within each group of models, the gap widens from 26\% on small puzzles to 37\% on large ones.
Notably, both proprietary models solve at least 50\% of the large puzzles, whereas none of the open-weight models solves more than 50\% even of the small puzzles.
Codoku thus sits at neither ceiling nor floor for current models:
our puzzles resolve differences between them.

\smalltitle{Larger profiles are generally harder}
Every model solves fewer large puzzles than small ones,
and the decline is monotonic for all models except GPT 5.6 Sol, which performs about equally on medium and large puzzles.
Averaged over the five models, the solve rate falls from 56\% (small) to 39\% (medium) and 30\% (large).
The decline is steeper for open-weight models, which retain about 30\% of their small-profile solve rate on large puzzles, against about 70\% for proprietary ones.

\begin{figure*}[tb]
    \centering
    \includegraphics[width=\textwidth]{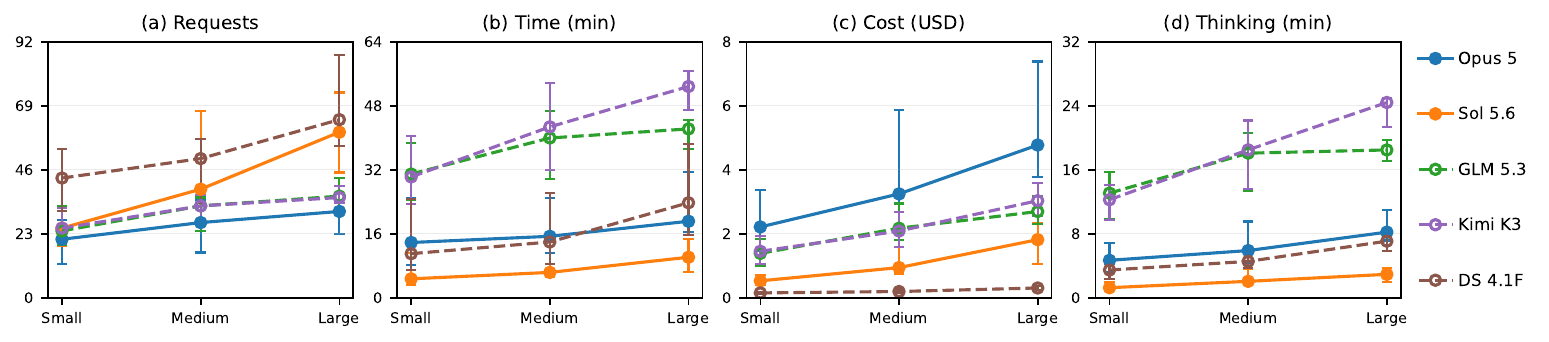}
    \caption{
        \textbf{Resource use rises with profile scale among solved runs}.
        Each plot reports the per-run median and interquartile range.
    }
    \label{fig:resource_usage}
\end{figure*}

\smalltitle{Larger profiles demand more resources}
Among solved runs, the median request count, wall-clock time, model cost in dollars, and thinking time all increase from small to medium to large for every model (\Cref{fig:resource_usage}).
Models trade these resources off differently.
GPT 5.6 Sol is the fastest on every profile, taking a median of 10.1 minutes per solved large puzzle against 19.1 for Claude Opus 5, although the former issues nearly twice as many requests.
DeepSeek V4.1 Flash is the cheapest on every profile yet issues the most requests, whereas Claude Opus 5 is the most expensive, at a median of \$4.77 per solved large puzzle.
GLM 5.3 and Kimi K3 are by far the slowest:
their median solved large puzzle takes 42.2 and 52.8 minutes, close to the one-hour budget.
This suggests that the time limit contributes to their high exhaustion rates on large puzzles;
indeed, the thinking-time plot shows that both models spend substantial time thinking.
A breakdown is presented in \Cref{sec:app_cost}.

\smalltitle{Answer to RQ2}
Codoku poses a substantial challenge to all five evaluated models under the Pi agent and our budgets:
none approaches saturation on any profile, every model solves fewer large puzzles than small ones, and solved runs demand more resources as the profiles grow.
The benchmark also separates proprietary from open-weight models, by a margin that widens with puzzle size.

\section{Discussion}\label{sec:discussion}

\subsection{Solving Strategies}\label{ssec:strategies}

\begin{figure*}[tb]
    \centering
    \includegraphics[width=\textwidth]{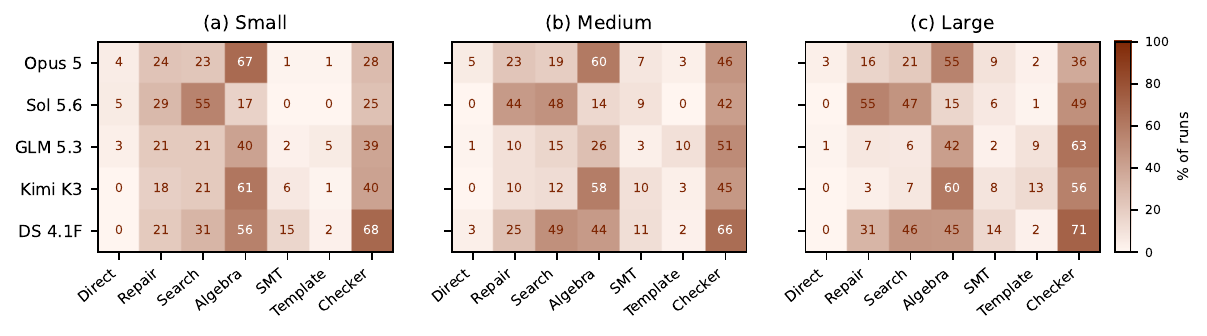}
    \caption{
        \textbf{Agents combine several strategies to solve codokus}.
        Each cell reports the percentage of solved runs that carry the corresponding strategy label.
        A run may carry several labels.
    }
    \label{fig:strategy_heatmap}
\end{figure*}

To characterize how agents solve codokus, we first inspected a sample of trajectories manually and identified seven recurring strategies (see \Cref{sec:app_case} for an example):
\begin{itemize}
    \item \code{Direct}: fill all cells in a single attempt;
    \item \code{Repair}: revise a candidate after a failed check and check the revision again;
    \item \code{Search}: enumerate or sample candidate fillings with an explicit script;
    \item \code{Algebra}: infer program states, inputs, or outputs algebraically from the constraints;
    \item \code{SMT}: encode local states or fillings as constraints and solve them with an SMT solver;
    \item \code{Template}: fill cells through a template of numbered slots and a mapping from slots to fillings;
    \item \code{Checker}: inspect the source code of the solution checker.
\end{itemize}
We then label the trajectories of all solved runs using keyword-based heuristics, counting only high-confidence labels (\Cref{fig:strategy_heatmap}).
For instance, a failed check followed by an edit and a recheck counts as evidence of \code{Repair},
and executing a Python script that uses \code{Z3} or \code{CVC5} as evidence of \code{SMT}.

\smalltitle{Agents combine strategies, and the most frequent one differs by model}
For every model and profile, a run carries more than one label on average.
No strategy is the most frequent for every model:
\code{Algebra} is the most frequent label for Claude Opus 5 and Kimi K3 on every profile,
\code{Checker} for DeepSeek V4.1 Flash on every profile,
\code{Search} or \code{Repair} for GPT 5.6 Sol,
and \code{Algebra} or \code{Checker} for GLM 5.3.
Every model carries each label on at least one profile, except Kimi K3, which never carries the \code{Direct} label.
The share of runs that inspect the checker's source code is higher on large than on small puzzles for every model.
Because the checker evaluates candidates against $\Phi$ rather than against the witness (\Cref{ssec:solution_verification}),
its source reveals how the constraints are checked rather than the fillings of the witness.
Agents typically begin solving a puzzle by reading the checker.

\smalltitle{Agents infer program states and confine their search}
\code{Direct} appears in at most 5\% of the runs of any model and profile, which suggests that agents treat the cells as interdependent.
Instead of filling all cells at once, runs derive cell values from the program states that the constraints entail,
either algebraically (\code{Algebra}, 14\%--67\% of runs) or with an SMT solver (\code{SMT}, up to 15\%).
Where agents search, in the trajectories we inspected, they confine the search to restricted domains,
such as the constants in the table or values around a ``pivot'' inferred from a local program state;
exhaustive enumeration is infeasible in any case, given the size of the search space (\Cref{ssec:rq1}).
When a candidate fails a check, agents also revise it and check it again (\code{Repair}, 3\%--55\% of runs).
Together, these observations indicate that agents solve codokus by reasoning about program states and narrowing their search accordingly, rather than by exhaustively enumerating fillings.

\subsection{Limitations and Threats to Validity}\label{ssec:limitations}

\smalltitle{Limitations}
Codokus are limited by dead blocks, since not every cell lies on the executed path.
A cell in a block that the execution never reaches is constrained only statically, and may therefore admit more fillings and be easier to fill;
in \Cref{fig:codoku-example}, \codokucell{ID} of the unexecuted block \code{b2} admits all three variables.
Input-output examples whose executions jointly cover every basic block would constrain such cells dynamically; we leave their generation to future work.

\smalltitle{Threats to validity}
Our results depend on the time, cost, and request limits.
784 unsuccessful runs end at a limit while the agent is still at work;
a detailed breakdown is presented in \Cref{sec:app_failures}.
We adopt limits in line with standard coding-agent setups for repository-level tasks~\citep{cybergym,exploitgym}, which we consider reasonable for the substantially smaller codoku puzzles.
Larger budgets may raise solve rates, but doubling the time (2h), cost (\$30), and request (256) budgets for GPT 5.6 Sol with the large profile yields only modest gains:
just eight additional cases are solved out of 54.
This suggests that codoku puzzles remain challenging even with substantially larger budgets.
The strategy analysis of \Cref{ssec:strategies} relies on keyword-based heuristics:
although we count only high-confidence labels, the heuristics may miss strategies that leave no lexical trace in a trajectory and mislabel others,
and our observations on how agents confine their search rest on manual inspection of a sample of trajectories.
Finally, codokus are synthetic programs, whose distribution may differ from that of human-written code.
We argue that an agent or model claimed to have the ability to reason about programs, however, should be able to reason about synthetic ones as well.

\section{Related Work}\label{sec:related_work}

Program-reasoning benchmarks commonly study the relation among a program, its inputs, and its execution.
Codoku differs from them both in how its tasks are specified and in how they are supplied.

\smalltitle{Execution and input prediction}
CruxEval~\citep{cruxeval} asks a model to predict the output of a short Python function on an input, or an input that yields a given output;
CruxEval-X~\citep{cruxevalx} extends it to multiple programming languages.
CodeI/O~\citep{codeio} uses both directions as training tasks.
REval~\citep{reval} extends prediction to intermediate states, statement coverage, and execution paths, CoRe~\citep{core} asks about control and data dependencies,
and further studies test prediction under code mutations and input perturbations~\citep{canllmsreason,robustexecution}.
These benchmarks remain informative for bare models, but each task comes with a complete program that an agent can execute or instrument instead of reasoning about it (\Cref{sec:introduction}).
In a pilot study, GLM 5.2 (OpenCode) solves all 1,600 CruxEval tasks, each within five minutes.

\smalltitle{Program synthesis}
Programming by example (PBE) asks a solver to construct a program consistent with given input-output examples.
\citet{pbellm} evaluate LLMs on classic PBE domains, including text-editing problems from SyGuS~\citep{crossbeam} and PROSE~\citep{prose},
and CodeARC~\citep{codearc} derives PBE tasks for LLM agents from 1,114 functions in HumanEval, MBPP, and APPS~\citep{humaneval,mbpp,apps}, allowing agents to query the hidden target function on new inputs.
Because correctness is defined as agreement with the target on all inputs, a checker can only approximate it:
comparison with a reference implementation rejects equivalent alternatives, whereas finite testing may accept incorrect programs.
CodeARC checks candidates by differential testing with inputs generated by Pynguin and Mokav~\citep{pynguin,mokav},
yet with the same model and agent under a five-minute limit per task, its success rate ranges from 57.4\% to 72.6\% depending on the strictness of the checker.
Codoku instead makes the specification explicit: correctness is defined by the global constraints of a puzzle alone,
so the checker decides it exactly, accepting every filling that satisfies them, including fillings that differ from the witness, and rejecting all others.

\section{Conclusion}\label{sec:conclusion}

We presented Codoku, a renewable benchmark that evaluates program reasoning in coding agents through constrained program synthesis.
Codoku generates fresh, controlled puzzles with known solutions and explicit semantic constraints.
Despite full tool access and compact puzzle sizes, current agents remain challenged, with the best solving 54\% of large puzzles.
Codoku complements existing benchmarks by providing scalable and automatically checked synthesis tasks for coding agents.

\bibliography{main}
\bibliographystyle{icml2026}

\newpage
\appendix
\onecolumn
\section{Appendix}

\subsection{Puzzle Complexity}\label{sec:app_complexity}

\Cref{tab:app_complexity} reports detailed statistics about the 300 puzzles that we use in our evaluation.
It extends \Cref{tab:puzzle_complexity} with more metrics.
Following the conclusion in \Cref{ssec:rq1}, all metrics except two increase from the small to the medium to the large profile.
The two exceptions are the density of the data-dependency graph, which decreases from 0.04 to 0.01,
and the number of loops, whose median is one in every profile;
the number of loop iterations, in contrast, increases across the profiles.

\begin{table}[tb]
    \centering
    \footnotesize
    \setlength{\tabcolsep}{.35em}
    \newcommand{\meanstd}[2]{#1 & #2}
    \newcommand{\quartiles}[3]{#1 & #2 & #3}
    \caption{
      \textbf{Puzzle metrics per profile.} Values are rounded to one decimal place.
      Each profile reports mean with its standard deviation and median with the first and third quartile.
    }
    \label{tab:app_complexity}
\begin{tabular}{lrrrrrrrrrrrrrrr}
\toprule
& \multicolumn{5}{c}{\bf Small} & \multicolumn{5}{c}{\bf Medium} & \multicolumn{5}{c}{\bf Large} \\
\cmidrule(lr){2-6} \cmidrule(lr){7-11} \cmidrule(lr){12-16}
\bf Metric & \it Mean & \it Std. & \it Med. & \it Q1 & \it Q3 & \it Mean & \it Std. & \it Med. & \it Q1 & \it Q3 & \it Mean & \it Std. & \it Med. & \it Q1 & \it Q3 \\
\midrule
\multicolumn{16}{c}{\textbf{\textit{\underline{search space}}}}
\vspace{2pt} \\
Typed cells & \meanstd{58.5}{17.4} & \quartiles{58.5}{46.0}{71.3} & \meanstd{134.7}{62.4} & \quartiles{114.5}{89.8}{168.0} & \meanstd{366.5}{162.1} & \quartiles{313.0}{263.0}{444.8} \\
~~\code{ID} & \meanstd{11.9}{4.4} & \quartiles{11.5}{9.0}{15.0} & \meanstd{32.6}{18.3} & \quartiles{26.0}{20.0}{41.3} & \meanstd{95.1}{43.0} & \quartiles{81.5}{65.0}{115.0} \\
~~\code{FUNC} & \meanstd{0.0}{0.0} & \quartiles{0.0}{0.0}{0.0} & \meanstd{0.8}{1.1} & \quartiles{0.0}{0.0}{1.0} & \meanstd{2.9}{2.8} & \quartiles{2.0}{1.0}{4.0} \\
~~\code{CONST} & \meanstd{14.7}{5.7} & \quartiles{14.0}{10.0}{19.0} & \meanstd{32.1}{12.7} & \quartiles{29.0}{23.8}{39.0} & \meanstd{78.6}{33.5} & \quartiles{69.5}{54.8}{95.3} \\
~~\code{OP} & \meanstd{31.2}{10.8} & \quartiles{31.0}{22.0}{40.3} & \meanstd{65.9}{31.0} & \quartiles{56.0}{44.0}{80.3} & \meanstd{177.7}{83.8} & \quartiles{153.0}{122.8}{209.3} \\
~~\code{CTRL} & \meanstd{0.4}{0.6} & \quartiles{0.0}{0.0}{1.0} & \meanstd{1.6}{1.1} & \quartiles{1.0}{1.0}{2.0} & \meanstd{3.6}{3.7} & \quartiles{3.0}{1.0}{4.3} \\
~~\code{LABEL} & \meanstd{0.3}{1.3} & \quartiles{0.0}{0.0}{0.0} & \meanstd{1.7}{2.8} & \quartiles{0.0}{0.0}{4.0} & \meanstd{8.6}{6.2} & \quartiles{8.0}{4.0}{12.0} \\
Constant table entries & \meanstd{13.4}{5.1} & \quartiles{13.0}{10.0}{18.0} & \meanstd{28.2}{11.0} & \quartiles{26.0}{20.8}{33.3} & \meanstd{66.9}{27.3} & \quartiles{60.0}{46.0}{84.3} \\
Log10 space & \meanstd{72.0}{22.7} & \quartiles{73.4}{55.5}{87.0} & \meanstd{182.3}{89.5} & \quartiles{152.6}{118.4}{226.9} & \meanstd{540.0}{253.4} & \quartiles{451.0}{378.8}{661.3} \\
\multicolumn{16}{c}{\textbf{\textit{\underline{static structure}}}}
\vspace{2pt} \\
Lines of code & \meanstd{38.9}{6.8} & \quartiles{38.0}{34.8}{41.3} & \meanstd{71.9}{13.2} & \quartiles{69.0}{63.8}{82.0} & \meanstd{133.5}{22.9} & \quartiles{131.0}{118.0}{147.8} \\
Distinct operators & \meanstd{22.3}{2.5} & \quartiles{23.0}{21.0}{24.0} & \meanstd{28.1}{3.0} & \quartiles{28.0}{26.0}{31.0} & \meanstd{33.2}{2.7} & \quartiles{33.0}{31.0}{35.0} \\
Distinct operands & \meanstd{36.0}{7.7} & \quartiles{35.0}{31.0}{41.0} & \meanstd{72.9}{16.5} & \quartiles{69.0}{60.8}{82.0} & \meanstd{136.4}{37.7} & \quartiles{129.0}{109.8}{152.8} \\
Total operators & \meanstd{140.3}{41.1} & \quartiles{136.0}{115.8}{162.0} & \meanstd{292.8}{99.5} & \quartiles{266.5}{215.5}{351.5} & \meanstd{708.5}{262.0} & \quartiles{656.5}{524.3}{816.5} \\
Total operands & \meanstd{121.4}{45.0} & \quartiles{113.5}{96.0}{134.3} & \meanstd{303.4}{93.3} & \quartiles{283.5}{226.8}{364.0} & \meanstd{703.9}{210.3} & \quartiles{663.5}{581.3}{783.0} \\
Halstead diff. & \meanstd{38.6}{15.1} & \quartiles{34.4}{28.0}{44.8} & \meanstd{58.3}{12.9} & \quartiles{56.7}{48.6}{66.3} & \meanstd{86.2}{14.4} & \quartiles{84.4}{75.1}{96.1} \\
CFG nodes & \meanstd{4.8}{0.7} & \quartiles{5.0}{4.0}{5.0} & \meanstd{6.6}{1.1} & \quartiles{7.0}{6.0}{8.0} & \meanstd{9.9}{1.3} & \quartiles{10.0}{9.0}{11.0} \\
CFG edges & \meanstd{5.5}{1.0} & \quartiles{5.0}{5.0}{6.0} & \meanstd{8.9}{1.8} & \quartiles{9.0}{7.8}{10.0} & \meanstd{14.6}{2.2} & \quartiles{15.0}{13.0}{16.0} \\
Cyclo. compl. & \meanstd{2.7}{0.7} & \quartiles{3.0}{2.0}{3.0} & \meanstd{4.3}{1.0} & \quartiles{4.0}{4.0}{5.0} & \meanstd{6.7}{1.3} & \quartiles{7.0}{6.0}{8.0} \\
Data dep. nodes & \meanstd{21.4}{5.7} & \quartiles{20.0}{18.0}{22.0} & \meanstd{44.0}{10.5} & \quartiles{41.5}{36.8}{51.0} & \meanstd{93.0}{19.3} & \quartiles{91.5}{79.0}{104.5} \\
Data dep. edges & \meanstd{16.7}{9.3} & \quartiles{15.0}{12.8}{16.0} & \meanstd{46.3}{15.9} & \quartiles{44.0}{33.8}{55.3} & \meanstd{111.7}{31.7} & \quartiles{109.5}{90.0}{130.3} \\
Data dep. degree & \meanstd{1.5}{0.3} & \quartiles{1.4}{1.3}{1.6} & \meanstd{2.1}{0.3} & \quartiles{2.0}{1.9}{2.3} & \meanstd{2.4}{0.3} & \quartiles{2.4}{2.2}{2.6} \\
Data dep. max degree & \meanstd{6.7}{2.5} & \quartiles{6.0}{5.0}{8.0} & \meanstd{10.6}{3.5} & \quartiles{10.0}{8.0}{12.0} & \meanstd{23.3}{6.3} & \quartiles{23.0}{18.0}{28.0} \\
Data dep. density (\%) & \meanstd{4.0}{1.0} & \quartiles{4.0}{3.0}{4.0} & \meanstd{2.0}{0.0} & \quartiles{2.0}{2.0}{3.0} & \meanstd{1.0}{0.0} & \quartiles{1.0}{1.0}{1.0} \\
\multicolumn{16}{c}{\textbf{\textit{\underline{dynamic behavior}}}}
\vspace{2pt} \\
EP length & \meanstd{5.9}{1.6} & \quartiles{6.0}{5.0}{7.0} & \meanstd{9.7}{3.0} & \quartiles{9.0}{7.0}{11.3} & \meanstd{15.3}{3.8} & \quartiles{15.0}{12.0}{17.0} \\
Unique blocks & \meanstd{4.7}{0.7} & \quartiles{5.0}{4.0}{5.0} & \meanstd{5.9}{1.1} & \quartiles{6.0}{5.0}{7.0} & \meanstd{7.1}{1.5} & \quartiles{7.0}{6.0}{8.0} \\
Repeated blocks & \meanstd{1.2}{1.3} & \quartiles{1.0}{0.0}{2.0} & \meanstd{3.8}{2.5} & \quartiles{3.0}{2.0}{5.0} & \meanstd{8.2}{3.7} & \quartiles{7.0}{6.0}{10.0} \\
Max visits per block & \meanstd{1.5}{0.5} & \quartiles{2.0}{1.0}{2.0} & \meanstd{2.5}{0.6} & \quartiles{2.0}{2.0}{3.0} & \meanstd{4.1}{1.0} & \quartiles{4.0}{3.0}{5.0} \\
Loops & \meanstd{0.7}{0.5} & \quartiles{1.0}{0.0}{1.0} & \meanstd{1.1}{0.3} & \quartiles{1.0}{1.0}{1.0} & \meanstd{1.1}{0.2} & \quartiles{1.0}{1.0}{1.0} \\
Loop iterations & \meanstd{1.1}{0.9} & \quartiles{1.0}{0.0}{2.0} & \meanstd{2.5}{1.0} & \quartiles{2.0}{2.0}{3.0} & \meanstd{4.1}{1.0} & \quartiles{4.0}{3.0}{5.0} \\
\multicolumn{16}{c}{\textbf{\textit{\underline{input-output examples}}}}
\vspace{2pt} \\
Examples & \meanstd{3.7}{0.8} & \quartiles{3.5}{3.0}{4.0} & \meanstd{6.3}{1.1} & \quartiles{6.0}{5.0}{7.0} & \meanstd{9.0}{0.8} & \quartiles{9.0}{8.0}{10.0} \\
\bottomrule
\end{tabular}
\end{table}

\subsection{Tool Calls}

\begin{figure}[tb]
    \centering
    \includegraphics[width=.9\textwidth]{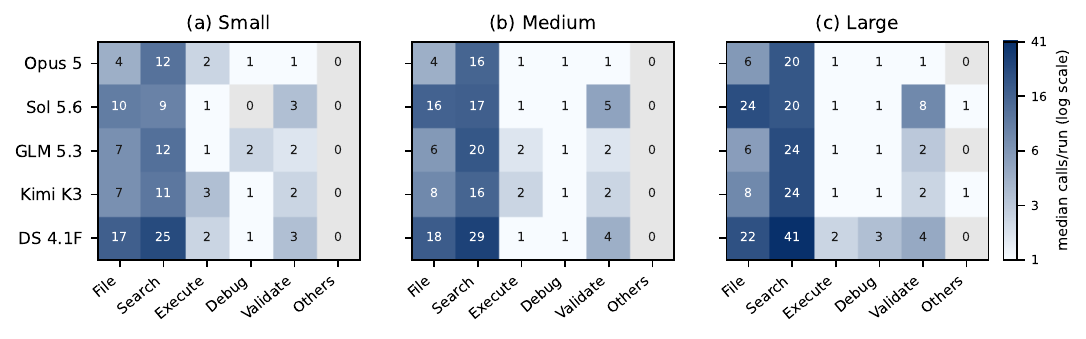}
    \caption{
        \textbf{Agents actively use tools, particularly for candidate search, but codoku puzzles make it difficult to translate tool calls into valid solutions}.
        \code{File}: file inspection and editing.
        \code{Search}: candidate search.
        \code{Execute}: candidate execution.
        \code{Debug}: debugging and instrumentation.
        \code{Validate}: syntax checking and solution verification.
        \code{Others}: all other tool calls.
    }
    \label{fig:toolcall_heatmap}
\end{figure}

We classify tool calls into six categories:
\code{File} for file inspection and editing,
\code{Validate} for syntax checking and solution verification,
\code{Execute} for candidate execution,
\code{Search} for candidate-search activity,
\code{Debug} for debugging and instrumentation, and \code{Others} for environment operations and remaining calls.
\Cref{fig:toolcall_heatmap} reports the median number of calls in each category per successful run, providing tool-level evidence that complements the strategy analysis in \Cref{ssec:strategies}.
Median calls categorized as candidate search increase across profiles for every model, while direct-execution calls remain between one and three, and debugging calls remain between zero and three.
GPT 5.6 Sol also makes more validation calls as profile scale increases.
This pattern is consistent with the increasing prevalence of \code{Repair} strategy in its successful trajectories.
DeepSeek V4.1 Flash makes the most tool calls overall on every profile, with medians increasing from 49 on small puzzles to 78.5 on large ones.
The tool-level \code{Search} category and the run-level \code{Search} strategy label should not be interpreted interchangeably.
The former counts operations categorized as candidate-search activity, whereas the latter requires high-confidence evidence of an explicit enumeration or sampling strategy.
Frequent search-category calls therefore do not imply that the explicit-search label appears in a comparable fraction of runs.

\subsection{Cost Breakdown}\label{sec:app_cost}

\Cref{tab:app_cost} breaks down the requests, tokens, and cost of every run, including solved and unsuccessful ones, and thus complements \Cref{fig:resource_usage}, which covers solved runs only.
Our experiment cost around \$4,000 in total:
$\sim$\$1,650 for Claude Opus 5, $\sim$\$575 for GPT 5.6 Sol, $\sim$\$665 for GLM 5.3, $\sim$\$1,030 for Kimi K3, and $\sim$\$85 for DeepSeek V4.1 Flash.
For every model, the mean number of requests, the mean number of tokens, and the mean cost per run increase from small to medium to large puzzles,
just as the median requests and cost increase among solved runs in \Cref{fig:resource_usage}.

\begin{table}[tb]
    \centering
    \footnotesize
    \caption{
        \textbf{Average requests, token usage, cost, and thinking time}, reported over all 1,500 agent runs.
    }
    \label{tab:app_cost}
\begin{tabular}{llrrrrrrrr}
\toprule
 & & & \multicolumn{5}{c}{\bf Tokens (K)} & & \\
\cmidrule(lr){4-8}
\bf Model & \bf Profile & \bf Requests & \it Input & \it Cache read & \it Cache write & \it Output & \it Think & \bf Cost (\$) & \bf Think (m) \\
\cmidrule(lr){1-10}
\multirow{3}{*}{\makecell[l]{\bf Claude\\Opus 5}}
  & \it $\cdot$~small & 22.8 & 0.0 & 1,494.5 & 105.7 & 67.5 & 56.9 & 3.49 & 6.6 \\
  & \it $\cdot$~medium & 29.7 & 0.1 & 2,312.4 & 283.9 & 97.2 & 84.6 & 5.85 & 9.4 \\
  & \it $\cdot$~large & 33.6 & 0.1 & 2,934.5 & 368.8 & 123.2 & 111.1 & 7.20 & 12.3 \\
\cmidrule(lr){1-10}
\multirow{3}{*}{\makecell[l]{\bf GPT\\5.6 Sol}}
  & \it $\cdot$~small & 39.7 & 0.1 & 1,126.1 & 39.9 & 19.9 & 10.2 & 1.20 & 2.2 \\
  & \it $\cdot$~medium & 60.0 & 0.6 & 2,349.9 & 56.8 & 24.8 & 14.6 & 1.75 & 3.3 \\
  & \it $\cdot$~large & 75.1 & 0.2 & 3,894.2 & 72.8 & 29.1 & 16.6 & 2.77 & 4.0 \\
\cmidrule(lr){1-10}
\multirow{3}{*}{\makecell[l]{\bf GLM\\5.3}}
  & \it $\cdot$~small & 21.8 & 245.1 & 2,009.6 & 0.0 & 124.3 & 116.5 & 1.68 & 14.7 \\
  & \it $\cdot$~medium & 23.5 & 307.2 & 2,478.7 & 0.0 & 156.1 & 149.0 & 2.20 & 19.0 \\
  & \it $\cdot$~large & 28.9 & 363.9 & 3,053.8 & 0.0 & 169.3 & 160.7 & 2.74 & 21.0 \\
\cmidrule(lr){1-10}
\multirow{3}{*}{\makecell[l]{\bf Kimi\\K3}}
  & \it $\cdot$~small & 26.8 & 31.2 & 1,585.8 & 113.0 & 79.1 & 67.5 & 2.58 & 15.2 \\
  & \it $\cdot$~medium & 29.8 & 42.1 & 2,008.4 & 136.0 & 96.9 & 85.8 & 3.36 & 19.2 \\
  & \it $\cdot$~large & 33.1 & 103.5 & 2,395.9 & 151.0 & 108.7 & 98.6 & 4.36 & 22.8 \\
\cmidrule(lr){1-10}
\multirow{3}{*}{\makecell[l]{\bf DPSK\\V4.1\\Flash}}
  & \it $\cdot$~small & 61.7 & 60.7 & 6,911.5 & 0.0 & 115.1 & 99.4 & 0.22 & 4.7 \\
  & \it $\cdot$~medium & 73.9 & 80.5 & 9,278.4 & 0.0 & 138.4 & 122.8 & 0.28 & 5.9 \\
  & \it $\cdot$~large & 84.6 & 97.5 & 12,179.5 & 0.0 & 154.3 & 136.3 & 0.32 & 6.7 \\
\cmidrule(lr){1-10}
\multicolumn{2}{l}{\bf All runs} & 43.0 & 88.9 & 3,735.2 & 88.4 & 100.3 & 88.7 & 2.66 & 11.1 \\
\bottomrule
\end{tabular}
\end{table}

\subsection{Failure Analysis}\label{sec:app_failures}

\smalltitle{Termination reasons}
\Cref{tab:app_termination} divides the exhausted and failed runs of \Cref{tab:agent_results} by the recorded reason for their termination.
Of the 784 exhausted runs, 694 reach the time limit, 56 the request limit, and 34 the cost limit.
Of the 92 failed runs, 63 end with the agent stopping without a valid solution, 17 with an infrastructure error, 8 with a safety refusal, and 4 with an output-length error.
The reasons differ across models.
All cost limits, safety refusals, and output-length errors occur in runs of Claude Opus 5,
and all request limits in runs of GPT 5.6 Sol and DeepSeek V4.1 Flash.

\begin{table}[tb]
    \centering
    \footnotesize
    \caption{
        \textbf{Termination reason of all agent runs}.
        The last row sums over all 1,500 runs.
        \emph{Exhausted} and \emph{Failed} group the reasons as in \Cref{tab:agent_results}.
    }
    \label{tab:app_termination}
\begin{tabular}{llrrrrrrrr}
\toprule
 & & & \multicolumn{3}{c}{\bf Exhausted} & \multicolumn{4}{c}{\bf Failed} \\
\cmidrule(lr){4-6}\cmidrule(lr){7-10}
\bf Model & \bf Profile & \bf Solved & \it Time & \it Cost & \it Requests & \it Stopped & \it Refusal & \it Length & \it Infra. \\
\cmidrule(lr){1-10}
\multirow{3}{*}{\makecell[l]{\bf Claude\\Opus 5}}
  & \it Small & 77 & 18 & 2 & 0 & 0 & 3 & 0 & 0 \\
  & \it Medium & 62 & 25 & 8 & 0 & 0 & 4 & 0 & 1 \\
  & \it Large & 50 & 21 & 24 & 0 & 0 & 1 & 4 & 0 \\
\cmidrule(lr){1-10}
\multirow{3}{*}{\makecell[l]{\bf GPT\\5.6 Sol}}
  & \it Small & 67 & 32 & 0 & 1 & 0 & 0 & 0 & 0 \\
  & \it Medium & 53 & 38 & 0 & 5 & 4 & 0 & 0 & 0 \\
  & \it Large & 54 & 24 & 0 & 9 & 12 & 0 & 0 & 1 \\
\cmidrule(lr){1-10}
\multirow{3}{*}{\makecell[l]{\bf GLM\\5.3}}
  & \it Small & 50 & 40 & 0 & 0 & 9 & 0 & 0 & 1 \\
  & \it Medium & 22 & 53 & 0 & 0 & 18 & 0 & 0 & 7 \\
  & \it Large & 12 & 80 & 0 & 0 & 5 & 0 & 0 & 3 \\
\cmidrule(lr){1-10}
\multirow{3}{*}{\makecell[l]{\bf Kimi\\K3}}
  & \it Small & 48 & 51 & 0 & 0 & 1 & 0 & 0 & 0 \\
  & \it Medium & 28 & 72 & 0 & 0 & 0 & 0 & 0 & 0 \\
  & \it Large & 11 & 89 & 0 & 0 & 0 & 0 & 0 & 0 \\
\cmidrule(lr){1-10}
\multirow{3}{*}{\makecell[l]{\bf DPSK\\V4.1\\Flash}}
  & \it Small & 39 & 49 & 0 & 12 & 0 & 0 & 0 & 0 \\
  & \it Medium & 29 & 53 & 0 & 12 & 3 & 0 & 0 & 3 \\
  & \it Large & 22 & 49 & 0 & 17 & 11 & 0 & 0 & 1 \\
\cmidrule(lr){1-10}
\multicolumn{2}{l}{\bf All runs} & 624 & 694 & 34 & 56 & 63 & 8 & 4 & 17 \\
\bottomrule
\end{tabular}
\end{table}

\smalltitle{Fine-grained checker verdicts}
\Cref{tab:app_verdicts} reports the solution checker's verdict on all agent runs.
Runs that end at a budget can still carry a final candidate solution to check:
all 89 unsuccessful runs of Kimi K3 on large puzzles reach the time limit, and 5 of them end with a candidate that fails a check.
Of the 876 unsuccessful runs, 815 end without a candidate solution.
The remaining 61 end with a candidate that fails a check:
26 at the execution path, 21 at re-masking, and 14 at the other checks combined.
No final candidate fails at compilation or at the constant check.
Across the runs, every solved run executes a candidate at least once, with a median of five executions per run,
whereas 17\% of the unsuccessful runs do so, with a median of zero.

\begin{table}[tb]
    \centering
    \footnotesize
    \caption{
        \textbf{Fine-grained checker verdict on the final candidate of each run}.
        \emph{No sol.}: no provided solution.
        The remaining columns give the first check that the candidate fails.
        \emph{Basics} checks if all cells are filled,
        \emph{Re-mask.} is the cell and structure preservation check, and
        \emph{Exec.\ limit} is the checker's execution time limit (5s) for the candidate (\Cref{ssec:solution_verification}).
    }
    \label{tab:app_verdicts}
\begin{tabular}{llrrrrrrrrrrr}
\toprule
\bf Model & \bf Profile & \makecell{\bf No\\\bf sol.} & \bf Basics & \bf Parse & \bf Compile & \makecell{\bf Re-\\\bf mask.} & \bf CFG & \makecell{\bf Exec.\\\bf limit} & \bf EP & \bf Output & \bf Const. & \bf Pass \\
\cmidrule(lr){1-13}
\multirow{3}{*}{\makecell[l]{\bf Claude\\Opus 5}}
  & \it $\cdot$~small  & 22 & 0 & 0 & 0 & 0 & 0 & 1 & 0 & 0 & 0 & 77 \\
  & \it $\cdot$~medium & 37 & 1 & 0 & 0 & 0 & 0 & 0 & 0 & 0 & 0 & 62 \\
  & \it $\cdot$~large  & 47 & 0 & 0 & 0 & 0 & 0 & 0 & 3 & 0 & 0 & 50 \\
\cmidrule(lr){1-13}
\multirow{3}{*}{\makecell[l]{\bf GPT\\5.6 Sol}}
  & \it $\cdot$~small  & 32 & 0 & 0 & 0 & 0 & 0 & 0 & 1 & 0 & 0 & 67 \\
  & \it $\cdot$~medium & 42 & 0 & 0 & 0 & 2 & 0 & 0 & 0 & 3 & 0 & 53 \\
  & \it $\cdot$~large  & 43 & 0 & 0 & 0 & 1 & 0 & 0 & 2 & 0 & 0 & 54 \\
\cmidrule(lr){1-13}
\multirow{3}{*}{\makecell[l]{\bf GLM\\5.3}}
  & \it $\cdot$~small  & 46 & 0 & 0 & 0 & 3 & 0 & 0 & 1 & 0 & 0 & 50 \\
  & \it $\cdot$~medium & 74 & 0 & 1 & 0 & 1 & 0 & 0 & 2 & 0 & 0 & 22 \\
  & \it $\cdot$~large  & 83 & 0 & 1 & 0 & 3 & 1 & 0 & 0 & 0 & 0 & 12 \\
\cmidrule(lr){1-13}
\multirow{3}{*}{\makecell[l]{\bf Kimi\\K3}}
  & \it $\cdot$~small  & 50 & 0 & 0 & 0 & 2 & 0 & 0 & 0 & 0 & 0 & 48 \\
  & \it $\cdot$~medium & 68 & 0 & 0 & 0 & 4 & 0 & 0 & 0 & 0 & 0 & 28 \\
  & \it $\cdot$~large  & 84 & 0 & 0 & 0 & 2 & 0 & 1 & 2 & 0 & 0 & 11 \\
\cmidrule(lr){1-13}
\multirow{3}{*}{\makecell[l]{\bf DPSK\\V4.1\\Flash}}
  & \it $\cdot$~small  & 59 & 1 & 1 & 0 & 0 & 0 & 0 & 0  & 0 & 0 & 39 \\
  & \it $\cdot$~medium & 65 & 0 & 0 & 0 & 2 & 0 & 0 & 4  & 0 & 0 & 29 \\
  & \it $\cdot$~large  & 63 & 1 & 1 & 0 & 1 & 1 & 0 & 11 & 0 & 0 & 22 \\
\cmidrule(lr){1-13}
\multicolumn{2}{l}{\bf All runs} & 815 & 3 & 4 & 0 & 21 & 2 & 2 & 26 & 3 & 0 & 624 \\
\bottomrule
\end{tabular}
\end{table}

\subsection{Budgets and Effort Scaling}\label{sec:app_scale}

We evaluate whether codokus remain challenging with larger resource budgets and higher reasoning effort.
We use the large profile and select GPT 5.6 Sol because it performs best on this profile (\Cref{tab:agent_results}), leading to 54 puzzles.
Budget constraints prevent us from conducting more experiments.

\smalltitle{Double budget}
We double the time, cost, and request budgets to 2 hours, \$30, and 256 requests.
This yields eight additional solved puzzles out of 54, while 40 puzzles still exceed the time limit.

\smalltitle{Xhigh effort}
We increase the reasoning effort to \code{xhigh}.
This yields 11 additional solved puzzles out of 54, while 23 puzzles still exceed the time limit.

\subsection{Case Study}\label{sec:app_case}

We analyze a small Codoku puzzle and how agents solve it.

\smalltitle{Puzzle}
The puzzle has five basic blocks (Lines 3--6).
Four of them, \ie, \code{entry}, \code{b0}, \code{b1}, and \code{b2}, form a loop (Lines 51--68), but the execution follows a straight-line path because the loop runs for only one iteration, as specified by the execution path (Line 15).
The constant table contains 13 entries with 18 total occurrences, with at most three occurrences per constant (Line 43).
The puzzle provides three input-output examples (Lines 73--75) that call \code{codoku}.
Four checksums constrain the filling:
the return value of each of the three example calls (Lines 73--75) and the accumulated checksum after the third call (Line 76).
This structure resembles sudoku puzzles, which imposes local constraints on each $3\times3$ block and global constraints across the $9\times9$ grid.
In codoku puzzles, the three return-value checksums act as local constraints, while the accumulated checksum acts as a global constraint across the examples.
The banner comment lists additional requirements that a valid solution must satisfy (Lines 29--32).
These features are consistent with the design described in \Cref{sec:codoku}.

\begin{figure}[tb]
    \centering
\renewcommand{\codefontsize}{\tiny}
\begin{lstlisting}
# codoku() is a function of the following CFG:
#
#@CFG_EDGE: entry -> b0, exit
#@CFG_EDGE: b0 -> b1
#@CFG_EDGE: b1 -> b2
#@CFG_EDGE: b2 -> entry, exit
#
# ------------------------------------------------
# Task
# ------------------------------------------------
#
# Replace all occurrences of <XXX> with appropriate code to make the function return the expected values for the examples
# in main following the below execution path:
#
#@EXE_PATH: entry -> b0 -> b1 -> b2 -> entry -> exit
#
# ------------------------------------------------
# Validation
# ------------------------------------------------
#
# Use the following command to verify your solution:
#
#   codoku check puzzle.py solution.py
#
# ------------------------------------------------
# General Requirements
# ------------------------------------------------
#
# 1. Each <XXX> cell must be filled out with a corresponding element.
# 2. You have access to all common command line tools and SMT solvers.
# 3. Do NOT change any code except for the <XXX> cells.
# 4. Do NOT introduce any new code, variables, or basic blocks.
#
# ------------------------------------------------
# Requirements for <CONST>
# ------------------------------------------------
#
# The line below list every constant the <CONST> cells must carry, as '<value>:<count>' pairs. Across your whole solution
# each <value> must appear in <CONST> positions exactly <count> times -- no more, no fewer -- and no other constant may
# appear in any <CONST> position. The value must match exactly, including its type: `2` (integer) and `2.0` (float) are
# distinct. Constants already shown in the fixed code do not count toward this budget.
#
#@CONST_TB: 1048577:1, 1946157169:3, 2:2, 2231:1, 24:1, 251:1, 32:1, 51:1, 62:1, 67108870:1, 71:1, 74261:1, 805306406:3
g__chk = [0]

def codoku(pa0, pa1, pa2):
  v0 = §\lstcodokucell{OP}§ §\lstcodokucell{CONST}§;  v1 = §\lstcodokucell{OP}§95;  v2 = -§\lstcodokucell{CONST}§;  t0 = [-§\lstcodokucell{CONST}§, _PAD, §\lstcodokucell{OP} §§\lstcodokucell{CONST}§, _PAD, _PAD];
  #@CFG_BLOCK entry
  t0[2] = ((cast_int(-3, 20) + cast_int(v1, 20)) - -524288)
  v2 = (cast_int(1912602622, 64) - (-13 * pa1))
  while (((1 §\lstcodokucell{OP}§ §\lstcodokucell{ID}§) §\lstcodokucell{OP}§ (§\lstcodokucell{OP}§1946157169 §\lstcodokucell{OP}§ pa0 §\lstcodokucell{OP}§ (-§\lstcodokucell{CONST}§ §\lstcodokucell{OP}§ pa0 §\lstcodokucell{OP}§ (§\lstcodokucell{OP}§ §\lstcodokucell{CONST}§ §\lstcodokucell{OP}§ 0)
          == (§\lstcodokucell{ID}§ §\lstcodokucell{OP}§ 0) §\lstcodokucell{OP}§ -(§\lstcodokucell{OP}§ §\lstcodokucell{OP}§ §\lstcodokucell{CONST}§ §\lstcodokucell{OP}§ §\lstcodokucell{ID}§))))) §\lstcodokucell{OP}§ ((§\lstcodokucell{OP}§ §\lstcodokucell{CONST}§ §\lstcodokucell{OP}§ §\lstcodokucell{ID}§)):
    #@CFG_BLOCK b0
    v1 = ((§\lstcodokucell{CONST}§ - §\lstcodokucell{ID}§ §\lstcodokucell{OP}§ (805306406 §\lstcodokucell{OP}§ §\lstcodokucell{ID}§ §\lstcodokucell{OP}§ (§\lstcodokucell{CONST}§ §\lstcodokucell{OP}§ 0) §\lstcodokucell{OP}§ (§\lstcodokucell{ID}§ < 0) §\lstcodokucell{OP}§ -(§\lstcodokucell{OP}§ §\lstcodokucell{CONST}§ §\lstcodokucell{OP}§ pa0)))
           - (§\lstcodokucell{OP}§ §\lstcodokucell{CONST}§ §\lstcodokucell{OP}§ §\lstcodokucell{ID}§))
    v0 = (cast_int(§\lstcodokucell{CONST}§, 8) §\lstcodokucell{OP}§ cast_int(v1, 8))
    #@CFG_BLOCK b1
    g__chk[0] = cast_int(global_chksum(g__chk[0], cast_int(v0, 32), v1, cast_int(v2, 32), ...), 64)
    v0 = (cast_int(§\lstcodokucell{CONST}§, 8) §\lstcodokucell{OP}§ cast_int(§\lstcodokucell{ID}§, 8))
    t0[0] = (pa2 §\lstcodokucell{OP}§ §\lstcodokucell{ID}§)
    #@CFG_BLOCK b2
    v2 = ((cast_int(§\lstcodokucell{OP}§ §\lstcodokucell{CONST}§, 64) §\lstcodokucell{OP}§ §\lstcodokucell{OP}§2147483647) - §\lstcodokucell{ID}§)
    v1 = (§\lstcodokucell{CONST}§ §\lstcodokucell{OP}§ (§\lstcodokucell{CONST}§ * pa0))
    if ((§\lstcodokucell{ID}§ §\lstcodokucell{OP}§ (§\lstcodokucell{OP}§32746 §\lstcodokucell{OP}§ §\lstcodokucell{ID}§))) §\lstcodokucell{OP}§ (((cast_int(§\lstcodokucell{OP}§1, 16) §\lstcodokucell{OP}§ 0) §\lstcodokucell{OP}§ (0 §\lstcodokucell{OP}§ (§\lstcodokucell{ID}§) §\lstcodokucell{OP}§ (0) §\lstcodokucell{OP}§ §\lstcodokucell{CONST}§))):
      §\lstcodokucell{CTRL}§
    #@CFG_BLOCK entry
    t0[2] = ((cast_int(-3, 20) + cast_int(v1, 20)) - -524288)
    v2 = (cast_int(1912602622, 64) - (-13 * pa1))
  #@CFG_BLOCK exit
  return global_chksum(0, cast_int(v0, 32), v1, cast_int(v2, 32), cast_int(_rd(t0, 0), 32), ...)

if __name__ == '__main__':
  r = codoku(-33554434, -4294967296, 22); check_chksum(-865773859, r)
  r = codoku(-546995, 576460752399126213, 11512); check_chksum(-667196848750, r)
  r = codoku(-131331, 0, -15130); check_chksum(-84085712, r)
  check_chksum(-667895151815, g__chk[0])
\end{lstlisting}
    \caption{
        \textbf{A small codoku puzzle example}.
        Claude Opus 5 solved it in $\sim$36 minutes while GPT 5.6 Sol, GLM 5.3, Kimi K3, and DeepSeek V4.1 Flash timed out with a one-hour budget.
    }
    \label{fig:app_case_puzzle}
\end{figure}

\smalltitle{Solution}
For this small puzzle, Claude Opus 5 solved it in $\sim$36 minutes while GPT 5.6 Sol, GLM 5.3, Kimi K3, and DeepSeek V4.1 Flash all timed out with a one-hour budget.
\Cref{fig:app_case_solution} displays Claude's solution.

\emph{Overview}.
Claude solves the puzzle by inspecting the checker, deriving checksum constraints, and combining enumeration with randomized search.
It then fills the remaining cells to satisfy the required control flow and constant table, assembles the solution, and passes the checker on its first attempt.

\emph{Phase 1: read the puzzle}.
From the puzzle, Claude infers that the loop runs once per call and the branch in \code{b2} (Lines 64--65) is not taken.
Because \code{v2} and \code{t0[2]} are fixed at Lines 67--68 before the \code{exit} block, the agent identifies \code{v0}, \code{v1}, and \code{t0[0]} as unknown exit values and decides to inspect the checker before searching.

\emph{Phase 2: inspect the checker}.
The agent reads the checker and its helper module to establish the filling rules.
Unary operators are allowed before constants, but \code{and} and \code{or} cannot fill \codokucell{OP} cells;
parameters and subscripts can fill \codokucell{ID} cells;
and the 18 \codokucell{CONST} cells must reproduce the constant table exactly.

\emph{Phase 3: reason local checksums}.
The agent simplifies the exit checksums using the known inputs inferred before and some shifts inferred by reading the \code{global\_chksum} function.
It reduces unknowns to \code{v1}, \code{t0[0]}, and the sign of \code{v0}.
Enumerating the \code{b1} and \code{b2} blocks with templates yields 824 matching combinations, all with \code{v1 = 67108870 + (2 * pa0)} and \code{t0[0]} equal to \code{0} in every call.
It later makes the latter happen using \code{t0[0] = pa2 - pa2}.

\emph{Phase 4: reason global checksum}.
Based on prior reasoning, the agent expresses the accumulated checksum as
$$
\textstyle\sum_{i=1}^{3} \big((y_i + A) \bmod m_i\big) - \sum_{i=1}^{3} K_i,
$$
where
$\bmod$ is Python's remainder,
$y_i$ is the intermediate value after the second shift in \code{global\_chksum},
$A = 7328\,c_3 - 1651\,c_4$ depends on the initial values, and
$m_i$ and $K_i$ depend on each call's inputs.
Using $y_1,y_3 \in \{0,-1\}$, it reduces the checksum to
$y_2 \equiv \rho \pmod{944{,}660{,}365}$.
It leverages the GNU \code{factor} utility to factor the candidate numbers.
This produces 17 candidate values for \code{v1}, but these appear difficult to realize with the \code{b0} template, so it turns to randomized search.

\emph{Phase 5: perform randomized search}.
The agent samples fillings of the \code{b0} template and filters them using modular constraints.
After \emph{96 million} unsuccessful samples, it verifies its algebra and finds that the search produces too few distinct values.
Restricting the operators improves diversity, and it yields a match within the allowed time budget.
The match fixes the \code{b0} expression and the initial values:
$\code{t0[0]} = -62$ and $\code{v1} = -95$.

\emph{Phase 6: fill remaining cells}.
Claude then fills the control-flow cells so that the loop runs once, the \code{b2} branch is not taken, and the \codokucell{CTRL} cell preserves the declared CFG (Lines 3--6).
It chooses assignments to \code{v0} that satisfy the checksum sign constraints (according to Phase 3) and distributes the remaining constants among checksum-irrelevant cells to match the table.

\emph{Phase 7: assemble and check}.
The agent assembles the 80 possible fillings with a Python script and executes them to confirm the required execution path for all three calls.
Then it checks them through the solution checker, which returns \code{PASS} for two fillings.
Claude submits one of them.

\begin{figure}[tb]
{\renewcommand{\codefontsize}{\tiny}
\begin{lstlisting}
def codoku(pa0, pa1, pa2):
  v0 = §\lstcodokucellsol{-}§ §\lstcodokucellsol{1946157169}§;  v1 = §\lstcodokucellsol{-}§95;  v2 = -§\lstcodokucellsol{1946157169}§;  t0 = [-§\lstcodokucellsol{62}§, _PAD, §\lstcodokucellsol{-}§ §\lstcodokucellsol{1946157169}§, _PAD, _PAD];
  #@CFG_BLOCK entry
  t0[2] = ((cast_int(-3, 20) + cast_int(v1, 20)) - -524288)
  v2 = (cast_int(1912602622, 64) - (-13 * pa1))
  while (((1 §\lstcodokucellsol{+}§ §\lstcodokucellsol{t0[0]}§) §\lstcodokucellsol{>}§ (§\lstcodokucellsol{-}§1946157169 §\lstcodokucellsol{*}§ pa0 §\lstcodokucellsol{*}§ (-§\lstcodokucellsol{2231}§ §\lstcodokucellsol{*}§ pa0 §\lstcodokucellsol{-}§ (§\lstcodokucellsol{-}§ §\lstcodokucellsol{24}§ §\lstcodokucellsol{*}§ 0)
          == (§\lstcodokucellsol{v0}§ §\lstcodokucellsol{*}§ 0) §\lstcodokucellsol{*}§ -(§\lstcodokucellsol{-}§ §\lstcodokucellsol{{\raisebox{.35ex}{\texttildelow}}}§ §\lstcodokucellsol{51}§ §\lstcodokucellsol{*}§ §\lstcodokucellsol{v1}§))))) §\lstcodokucellsol{<}§ ((§\lstcodokucellsol{-}§ §\lstcodokucellsol{805306406}§ §\lstcodokucellsol{<}§ §\lstcodokucellsol{pa2}§)):
    #@CFG_BLOCK b0
    v1 = ((§\lstcodokucellsol{805306406}§ - §\lstcodokucellsol{pa0}§ §\lstcodokucellsol{*}§ (805306406 §\lstcodokucellsol{//}§ §\lstcodokucellsol{pa0}§ §\lstcodokucellsol{-}§ (§\lstcodokucellsol{32}§ §\lstcodokucellsol{-}§ 0) §\lstcodokucellsol{+}§ (§\lstcodokucellsol{pa0}§ < 0) §\lstcodokucellsol{\%}§ -(§\lstcodokucellsol{+}§ §\lstcodokucellsol{1048577}§ §\lstcodokucellsol{//}§ pa0)))
           - (§\lstcodokucellsol{not}§ §\lstcodokucellsol{74261}§ §\lstcodokucellsol{//}§ §\lstcodokucellsol{pa2}§))
    v0 = (cast_int(§\lstcodokucellsol{2}§, 8) §\lstcodokucellsol{+}§ cast_int(v1, 8))
    #@CFG_BLOCK b1
    g__chk[0] = cast_int(global_chksum(g__chk[0], cast_int(v0, 32), v1, cast_int(v2, 32), ...), 64)
    v0 = (cast_int(§\lstcodokucellsol{251}§, 8) §\lstcodokucellsol{+}§ cast_int(§\lstcodokucellsol{t0[0]}§, 8))
    t0[0] = (pa2 §\lstcodokucellsol{-}§ §\lstcodokucellsol{pa2}§)
    #@CFG_BLOCK b2
    v2 = ((cast_int(§\lstcodokucellsol{-}§ §\lstcodokucellsol{71}§, 64) §\lstcodokucellsol{*}§ §\lstcodokucellsol{-}§2147483647) - §\lstcodokucellsol{t0[0]}§)
    v1 = (§\lstcodokucellsol{67108870}§ §\lstcodokucellsol{+}§ (§\lstcodokucellsol{2}§ * pa0))
    if ((§\lstcodokucellsol{t0[0]}§ §\lstcodokucellsol{+}§ (§\lstcodokucellsol{-}§32746 §\lstcodokucellsol{*}§ §\lstcodokucellsol{pa2}§))) §\lstcodokucellsol{*}§ (((cast_int(§\lstcodokucellsol{-}§1, 16) §\lstcodokucellsol{*}§ 0) §\lstcodokucellsol{*}§ (0 §\lstcodokucellsol{+}§ (§\lstcodokucellsol{t0[0]}§) §\lstcodokucellsol{+}§ (0) §\lstcodokucellsol{+}§ §\lstcodokucellsol{805306406}§))):
      §\lstcodokucellsol{break}§
    #@CFG_BLOCK entry
    t0[2] = ((cast_int(-3, 20) + cast_int(v1, 20)) - -524288)
    v2 = (cast_int(1912602622, 64) - (-13 * pa1))
  #@CFG_BLOCK exit
  return global_chksum(0, cast_int(v0, 32), v1, cast_int(v2, 32), cast_int(_rd(t0, 0), 32), ...)
\end{lstlisting}
}
    \caption{
        \textbf{Filling submitted by Claude Opus 5} for the puzzle of \Cref{fig:app_case_puzzle}.
        Highlighted tokens fill the cells; the remaining code is fixed.
        The filling is valid.
    }
    \label{fig:app_case_solution}
\end{figure}

\smalltitle{Observations}
Consistent with \Cref{ssec:strategies}, the agent
inspects the checker source (\code{Checker}),
algebraically derives constraints on intermediate values (\code{Algebra}),
searches for fillings through scripted enumeration and sampling (\code{Search}),
and fills the cells using a numbered list of slots (\code{Template}).
It does not use an SMT solver in this run, but invokes GNU \code{factor} to compute prime factorizations.
The run does not include \code{Repair}, as two of the 80 initial fillings already pass solution verification.
Phases 4 and 5 consume 25.8 of the run's 36.4 minutes and \$3.70 of its \$6.07 cost.
Some fillings reduce subexpressions to constants, as in \code{t0[0] = pa2 - pa2} and the zero factor in the \code{b2} condition;
these simplifications help Claude find a solution more quickly.
The checker accepts these fillings because it verifies the global constraints rather than requiring agreement with the witness (\Cref{ssec:solution_verification}).


\end{document}